\documentclass[11pt,a4paper]{article}
\usepackage[T1]{fontenc}
\usepackage[utf8]{inputenc}
\usepackage{lmodern,microtype}
\usepackage[margin=2.5cm]{geometry}
\usepackage{setspace}
\usepackage{amsmath,amssymb,booktabs,array,longtable,tabularx,pdflscape,graphicx,float,subcaption,tikz}
\usetikzlibrary{arrows.meta,positioning}
\usepackage[authoryear,round]{natbib}
\usepackage[hidelinks]{hyperref}
\setcitestyle{aysep={}}
\begin{document}
\begin{center}
{\Large\bfseries Validating LLM-Generated Synthetic Survey Samples:\\[3pt] A Use-Dependent Multi-Dimensional Framework for Social-Science Applications\par}
\vspace{18pt}
{\small Enrico Angelo Raffaele D'Ecclesiis$^{1,*}$\quad Fulvio Castellacci$^{2}$\quad Arash Ahmadi$^{3}$\par}
\vspace{13pt}
{\footnotesize
$^{1}$ Department of Economics ``Marco Biagi'', University of Modena and Reggio Emilia, Italy.\\
$^{*}$ Corresponding author: \href{mailto:enricoangeloraffaele.decclesiis@unimore.it}{enricoangeloraffaele.decclesiis@unimore.it}\\[6pt]
$^{2}$ TIK Centre for Technology, Innovation and Culture, Faculty of Social Sciences, University of Oslo, Norway.\\[6pt]
$^{3}$ Centre for Computational and Data Science (dScience), Faculty of Mathematics and Natural Sciences, University of Oslo, Norway.\par}
\vspace{14pt}
September 2026
\end{center}
\begin{abstract}

Large language models (LLMs) make it increasingly feasible to generate synthetic survey respondents, but their empirical usefulness cannot be judged by a single notion of accuracy. A synthetic sample may perform weakly at the respondent level while still preserving distributional or substantive relationships relevant to social-science analysis. We develop a use-dependent, multi-dimensional validation framework distinguishing predictive, distributional, and substantive validity, and illustrate it using Eurobarometer data on attitudes toward artificial intelligence. Four off-the-shelf LLMs generate responses conditional on observed respondent profiles, while supervised machine learning provides a predictive reference for interpreting their performance. The results reveal substantial differences across validation dimensions: some LLM-generated samples preserve meaningful marginal heterogeneity, demographic ordering, cross-country variation, and predictor structure despite imperfect respondent-level correspondence. Performance varies across models and outcomes, but the broader picture is cautiously constructive. Synthetic samples should therefore be evaluated against the empirical properties required by their intended use.
\end{abstract}
\clearpage
\section{Introduction}
\label{sec:introduction}

Large language models (LLMs) are increasingly being explored as synthetic respondents, simulated agents, and proxies for human participants in social-science research \citep{bail2024generative,ziems2024css}. By conditioning a model on a description of an individual and asking it to answer survey questions, researchers can construct respondent-level synthetic samples at relatively low cost \citep{argyle2023outofone,aher2023using,horton2023homo}. Such samples may be useful for questionnaire pre-testing, exploratory analysis, scenario simulation, and other settings in which conventional data collection is costly or time-consuming. Their growing use, however, raises a basic methodological question: when should an LLM-generated survey sample be considered empirically useful?

Plausible responses are not sufficient. A synthetic sample may reproduce an average response while failing to preserve individual variation, differences across social groups, or relationships between outcomes and respondent characteristics. Conversely, it may reproduce some aggregate patterns while showing only weak correspondence with the responses of the individuals whose profiles were supplied to the model. Recent studies document related concerns about respondent-level reliability, demographic representation, reproducibility, and the preservation of social heterogeneity \citep{bisbee2024synthetic,dominguez2024questioning,wang2024flatten,santurkar2023opinions,qu2024performance}. Existing evaluations also emphasize different empirical criteria, making synthetic samples appear more or less successful depending on the property being examined \citep{miranda2025simulating,yang2024large,sinacola2025llms}.

This paper proposes a use-dependent, multi-dimensional framework for validating LLM-generated synthetic survey samples in social-science applications. Existing research already evaluates synthetic respondents along multiple criteria, but these criteria are not generally organized according to the empirical purposes for which the resulting synthetic data are intended to be used. Our framework starts from this use-dependent perspective and distinguishes three complementary validation objects. \emph{Predictive validity} evaluates correspondence between generated and observed outcomes for the same respondents. \emph{Distributional validity} evaluates whether the synthetic sample reproduces the location, dispersion, and shape of the observed outcome distribution. \emph{Substantive validity} evaluates whether synthetic outcomes preserve empirical relationships relevant to downstream analysis, such as differences across demographic groups, geographic patterns, and predictor structure. These dimensions are not interchangeable: respondent-level accuracy, realistic marginal heterogeneity, and preservation of substantive relationships can diverge within the same synthetic sample.

The relevance of the three dimensions therefore depends on intended use. Respondent-level prediction or imputation places particular weight on predictive validity; applications intended to preserve marginal distributions and heterogeneity require distributional validity; and applications concerned with group differences or relationships between outcomes and respondent characteristics require substantive validity. We do not combine these dimensions into a universal validation score, since doing so would require arbitrary weights and could conceal trade-offs among empirically distinct properties. Instead, the framework provides a common workflow for identifying which empirical properties need to be validated for a given use and for assessing them separately.

We present an empirical illustration of this framework on the topic of individuals' attitudes towards AI. The application uses data from the Special Eurobarometer 554, \emph{Artificial Intelligence and the Future of Work}, fielded as part of Eurobarometer wave 101.4 across the European Union in 2024. The application focuses on three composite outcomes: (i) \emph{General}, capturing broad attitudes toward artificial intelligence and recent digital technologies; (ii) \emph{Effects}, capturing perceived effects and usefulness of AI in working life; and (iii) \emph{Use}, capturing attitudes toward concrete workplace applications of AI. Rather than asking the LLMs to generate a population from scratch, we condition them on profiles constructed from observed Eurobarometer respondents. For each respondent, the model receives a structured profile and the relevant questionnaire items and generates item-level responses, which are aggregated using the same rules as the observed survey data. The exercise therefore evaluates generated outcomes within the observed respondent-profile space; it does not test whether an LLM can independently reproduce the composition or representativeness of the European population.

We use four off-the-shelf LLMs from different providers: Claude Sonnet 4.6, GPT-4.1 mini, Gemini 3 Flash, and Grok 4.1 Fast Non-Reasoning. We use these LLMs without task-specific training, fine-tuning, or supervised calibration on the target survey outcomes. Their synthetic responses are compared with the observed outcomes. Further, we also carry out a comparison between observed outcomes and out-of-fold predictions from four conventional supervised learning methods: Elastic Net, Random Forest, Histogram Gradient Boosting, and Super Learner. The supervised ML models serve as a reference rather than as technically equivalent competitors. They learn directly from observed target outcomes, whereas the LLMs generate responses from respondent profiles and questionnaire information without target-specific supervised training. This asymmetry makes the benchmark informative about what can be recovered from the available respondent information when observed targets are available for learning.

The empirical results illustrate why a multi-dimensional assessment is necessary. The supervised models provide the strongest respondent-level predictions and reproduce observed means closely, but their predictions are substantially less dispersed than the observed outcomes. LLM performance is more heterogeneous. Claude provides the lowest respondent-level prediction error among the four LLMs, whereas other models sometimes reproduce the marginal dispersion and broader distributional shape of some of the outcome variables more closely. The exercise shows that greater synthetic dispersion is not uniformly beneficial, however, because it can coexist with sizeable mean bias or weak respondent-level correspondence.

The substantive-validity exercises provide a further perspective. In this Eurobarometer application, several LLM-generated samples preserve much of the observed ordering across age and education groups for the broader AI-attitude outcomes. Some models also recover meaningful cross-country variation and a substantial part of the predictor-ranking structure obtained from the observed data. These relationships are nevertheless model- and outcome-dependent. Attitudes toward concrete workplace uses of AI are particularly difficult to reproduce, and synthetic samples may preserve an ordering or predictor structure while remaining poorly calibrated in levels. We therefore interpret the findings as evidence of partial and dimension-specific agreement rather than as general validation of LLM-generated survey data.

The paper contributes to the literature in three related ways. First, it provides a use-dependent organization of existing and complementary validation criteria, linking paired respondent-level correspondence, marginal distributional fidelity, and preservation of analysis-relevant relationships to the empirical purposes for which synthetic survey data may be used. The contribution therefore lies in the explicit mapping between intended use and validation object rather than in multidimensional evaluation per se. Second, it clarifies the role of supervised machine learning as a reference for interpreting both predictive performance and distributional compression without treating ML and LLM outputs as equivalent technologies. Third, the multi-country application implements the three validation objects jointly for the same respondent profiles, outcomes, and synthetic samples across multiple LLMs, showing how the assessment of a synthetic sample changes with the empirical property required by the intended use.

The scope of these conclusions is deliberately limited. The empirical evidence comes from one survey, one substantive domain, and a specific set of models, respondent profiles, prompts, index definitions, and generation settings. The application therefore does not establish the general validity of synthetic survey respondents. Its broader contribution is methodological: LLM-generated survey samples should be treated as validation-dependent empirical objects, whose suitability must be established in relation to the properties required by the intended research use.

The remainder of the paper reviews the related literature, develops the validation framework, describes the Eurobarometer application and supervised benchmarks, and then reports predictive, distributional, and substantive validity before discussing implications, limitations, and conclusions.

\section{Related Literature and Positioning}
\label{sec:literature}

\subsection{Synthetic respondents and validation challenges}
\label{subsec:synthetic_respondents}

This paper contributes to the growing literature on large language models as synthetic respondents, simulated agents, and proxies for human participants. A foundational contribution is the notion of \emph{algorithmic fidelity} developed by \citet{argyle2023outofone}, which asks whether language models conditioned on socio-demographic profiles can reproduce response patterns associated with corresponding human subpopulations. Survey-based applications have extended this logic to domain-specific public opinion, showing that fidelity can depend on demographic and issue-relevant conditioning information \citep{lee2024climate}. Related work uses LLMs to simulate participants in behavioral and experimental settings \citep{aher2023using,horton2023homo} and has contributed to a broader debate over the conditions under which language models can serve as proxies for human subjects \citep{dillion2023replace}.

The attraction of these approaches is straightforward: LLMs can generate responses for alternative respondent profiles or hypothetical settings at relatively low marginal cost. Yet plausible model outputs do not necessarily constitute valid synthetic data. Recent studies document limitations in respondent-level reliability, demographic representation, and the preservation of heterogeneity. \citet{bisbee2024synthetic}, for example, show that synthetic responses can reproduce some aggregate properties while differing from observed survey data in variation and regression relationships. \citet{dominguez2024questioning} demonstrate that apparent demographic alignment can be sensitive to prompting choices, while \citet{wang2024flatten} show that LLM-based replacements can attenuate or distort differences across identity groups. Related persona-based evaluations likewise show that richer profile conditioning does not necessarily improve fidelity uniformly \citep{kaur2025syntheticvoices}.

A further challenge is that generated responses depend on the particular model, prompt, model version, and generation settings. The empirical properties of a synthetic sample therefore cannot be inferred from the general capabilities of an LLM but must be established for the specific generation procedure and application under consideration \citep{madden2025synthetic,abdurahman2025primer}. Taken together, this literature motivates treating LLM-generated respondents as empirical objects requiring explicit validation rather than as direct substitutes for observed human responses.

\subsection{Existing validation strategies and our positioning}
\label{subsec:positioning}

Existing studies validate synthetic respondents using heterogeneous criteria. Some approaches are explicitly multidimensional. \citet{argyle2023outofone}, for example, distinguish several components of algorithmic fidelity, including individual- and subgroup-level correspondence and preservation of empirical patterns. \citet{aher2023using} use ``Turing Experiments'' to ask whether simulated samples reproduce established results from human-subject studies. Other contributions focus more directly on statistical properties of observed and generated data. \citet{bisbee2024synthetic} compare means, variation, and regression relationships, while \citet{dominguez2024questioning} examine distributional and demographic alignment and its sensitivity to response ordering and labeling.

A further set of studies evaluates LLM responses primarily as predictions. \citet{miranda2025simulating} distinguish respondent-level predictive performance from distributional fit and compare LLM simulations with a Random Forest benchmark. \citet{yang2024large} examine social prediction under different information settings, showing that performance can depend strongly on the features supplied to the model. \citet{sinacola2025llms} likewise compare LLM-based and Random Forest predictions while examining heterogeneity in predictive performance across population groups. Table~A1 in the online supplement summarizes the validation strategies used in the studies most closely related to our framework.

Taken together, these strands of literature show that validation is already a central concern and that multidimensional evaluation is not itself new. This concern has become particularly prominent in recent methodological work on generative AI in the social sciences. The special issue introduced by \citet{davidson2025integrating}, for example, identifies rigorous measurement and validation of AI-generated outputs as one of the central methodological challenges raised by generative AI. Within the same issue, \citet{kozlowski2025simulating} emphasize the continued need for validation against human-subject data, \citet{lyman2025balancing} show that model choice and alignment can materially affect algorithmic fidelity, and \citet{boelaert2025machinebias} document systematic bias and restricted response variation when LLMs answer opinion-poll questions. Together, these contributions reinforce the view that the empirical properties of synthetic respondents must be established rather than inferred from the general capabilities of an LLM.

Recent work has also begun to systematize this validation problem in different ways. \citet{lin2026psychological} organizes validation primarily around the availability of human benchmark evidence, distinguishing direct, indirect, and generative validation, while \citet{xie2026statistical} focus on population-level statistical realism across distributions, associations, multivariate relationships, and longitudinal patterns. \citet{park2026llms} explicitly propose a multi-dimensional validation framework for synthetic innovation-survey responses, combining qualitative consistency, quantitative pattern replication, explanation quality, and robustness to generation choices. Complementing these approaches, \citet{cummins2026threat} shows that alternative silicon-sample configurations can differ substantially in their ability to reproduce respondent rankings, response distributions, and relationships between variables, and that strong performance on one criterion need not imply strong performance on another. Our framework therefore does not claim multidimensional validation itself as novel. Instead, it starts from the intended empirical use of a synthetic survey sample and distinguishes three corresponding validation objects: paired respondent-level correspondence (\emph{predictive validity}), properties of the marginal outcome distribution (\emph{distributional validity}), and preservation of analysis-relevant relationships (\emph{substantive validity}). Its contribution is to link these distinct objects explicitly to intended survey uses within a common empirical workflow, without assigning universal priority to any one dimension.

The paper also incorporates conventional supervised machine learning as a reference for interpreting the LLM-generated outcomes. This comparison is not intended as a horse race between technically equivalent methods: supervised models learn directly from observed target outcomes, whereas the LLMs in our application do not. Rather, supervised ML provides an established predictive reference for what can be recovered from the available respondent information when target data are available for learning \citep{mullainathan2017machine,athey2019machine}. The statistical role and interpretation of this benchmark are developed formally in the Validation Framework section.

Finally, our empirical application places particular emphasis on substantive validity. In addition to respondent-level and distributional correspondence, we ask whether synthetic outcomes preserve demographic differences, cross-country variation, and predictor structure that would matter in downstream social-science analysis. This aspect is paramount, given the crucial importance of empirical relationships among variables in social-science research. The objective here is not to establish that a particular LLM is generally valid, but to determine which empirical properties of an observed survey are reproduced by a particular synthetic sample and which are not, consistent with recent calls to assess the success and failure of LLMs relative to specific scientific tasks and uses \citep{argyle2025fickle}.

\section{Validation Framework}
\label{sec:framework}

\subsection{Validation objects and dimensions}
\label{subsec:framework_overview}

Let $i=1,\ldots,N$ index respondents, $Y_i$ denote an observed survey outcome, and $X_i$ the respondent characteristics used to construct the corresponding profile. For an LLM $m$, let $\widetilde{Y}_i^{(m)}$ denote the synthetic outcome obtained from its item-level responses and processed using the same coding and aggregation rules as the observed outcome. Observed and synthetic outcomes therefore refer to the same respondent profiles. The framework evaluates generated outcomes within this observed profile space rather than asking the LLM to generate the distribution of respondent characteristics themselves.

Validation concerns three distinct empirical objects: correspondence between observed and generated outcomes for the same respondents, correspondence between their marginal distributions, and preservation of relationships between outcomes and respondent characteristics. These define predictive, distributional, and substantive validity, respectively. The dimensions are complementary rather than hierarchical: agreement on one does not imply agreement on the others.

\subsubsection{Predictive validity}
\label{subsubsec:predictive_definition}

Predictive validity evaluates respondent-level correspondence. For model $m$, a general prediction loss is

\begin{equation}
D_{\mathrm{P}}^{(m)}
=
\frac{1}{N}
\sum_{i=1}^{N}
L\left(Y_i,\widetilde{Y}_i^{(m)}\right),
\label{eq:predictive_validity}
\end{equation}

where the loss function depends on the outcome and application. For the approximately continuous indices used here, we report MSE and RMSE and complement them with Pearson correlation. MSE is the squared-error criterion used to quantify prediction loss, while RMSE expresses the same error on the original outcome scale and is therefore easier to interpret substantively. Pearson correlation captures a different property: the extent to which generated outcomes reproduce systematic respondent-level covariation with the observed responses, independently of mean calibration. The three measures are therefore partly related but serve distinct interpretive purposes. Predictive validity is most directly relevant to applications involving respondent-level prediction, approximation, or imputation.

\subsubsection{Distributional validity}
\label{subsubsec:distributional_definition}

Distributional validity evaluates whether the marginal distribution of $\widetilde{Y}^{(m)}$ reproduces that of $Y$ in location, dispersion, and shape. No single statistic captures all three properties, and the diagnostics used here are deliberately complementary. Mean bias captures differences in location, while the ratio of synthetic to observed standard deviations captures relative dispersion. The Kolmogorov--Smirnov distance measures the largest discrepancy between the empirical cumulative distributions, whereas the Wasserstein distance summarizes the overall displacement required to align them. Errors at selected quantiles provide additional information on discrepancies at specific parts of the distribution, including its tails. These diagnostics overlap in the broad property they assess but are not interchangeable, and we complement them with graphical comparisons. For applications in which the synthetic sample is expected to preserve the marginal distribution and heterogeneity of survey responses, these distributional properties are central. Importantly, greater dispersion is not necessarily better: additional variation can reflect meaningful heterogeneity, but also bias, miscalibration, or model-specific noise.

\subsubsection{Substantive validity}
\label{subsubsec:substantive_definition}

Substantive validity asks whether synthetic outcomes preserve empirical relationships relevant to the intended analysis. Let $A(Y,X)$ denote an analysis performed using the observed outcome and respondent characteristics. The relevant comparison is

\begin{equation}
A(Y,X)
\;\approx\;
A\left(\widetilde{Y}^{(m)},X\right).
\label{eq:substantive_validity}
\end{equation}

The object $A(\cdot)$ is application-specific and can represent group comparisons, geographic patterns, regression relationships, predictor rankings, or other quantities relevant to the research question. In our application, we examine age and education patterns, cross-country variation, and predictor-ranking similarity. Substantive validity is therefore evaluated directly rather than inferred from respondent-level or distributional agreement. It is particularly relevant when the intended use of the synthetic sample depends on preserving these empirical relationships.

Table~\ref{tab:validation_framework} summarizes the three dimensions and their use-dependent interpretation.

\begin{table*}[t]
\small\sf\centering
\caption{Validation dimensions and their empirical use}
\label{tab:validation_framework}

\renewcommand{\arraystretch}{1.22}
\setlength{\tabcolsep}{4pt}

\begin{tabularx}{\textwidth}{
    p{1.9cm}
    X
    X
    X
}
\toprule
\textbf{Dimension} &
\textbf{Validation object} &
\textbf{Diagnostics in this application} &
\textbf{Most relevant uses} \\
\midrule

Predictive
&
Paired respondent-level outcomes
&
MSE, RMSE, Pearson correlation
&
Respondent-level prediction, approximation, or imputation
\\

\addlinespace

Distributional
&
Marginal outcome distribution
&
Mean bias, SD ratio, Kolmogorov--Smirnov and Wasserstein distances, quantile errors, distribution plots
&
Applications requiring realistic distributional heterogeneity
\\

\addlinespace

Substantive
&
Results of the same substantive analysis applied to observed and synthetic outcomes
&
Age- and education-group patterns, country-level variation, predictor-ranking similarity
&
Group comparisons, geographic analysis, and analysis of substantive relationships
\\

\bottomrule
\end{tabularx}

\end{table*}

The framework is consequently use-dependent rather than based on a universal validation score. Strong performance on one dimension does not compensate mechanically for weak performance on another when the latter is required by the intended use.

\subsection{Supervised machine learning as a reference}
\label{subsec:ml_benchmark_framework}

Observed survey outcomes remain the empirical reference throughout the framework. Besides comparing observed and synthetic responses, the framework incorporates a second comparison between observed responses and predictions generated by supervised machine-learning models. This second comparison plays a different role: it provides a credible reference for what can be predicted from the respondent information when observed target outcomes are available for learning. Supervised ML is an established predictive paradigm in applied social-science research \citep{mullainathan2017machine,athey2019machine}, whereas generating survey outcomes with prompt-conditioned general-purpose LLMs is a comparatively new empirical practice whose validity must be established rather than assumed.

The comparison is deliberately asymmetric. A supervised model learns a mapping from observed pairs $(X_i,Y_i)$, whereas the LLM receives a representation of $X_i$, questionnaire information, and generation instructions without task-specific supervised training or fine-tuning on the target outcomes. The benchmark is therefore not a horse race between equivalent methods. It indicates how much respondent-level structure is recoverable when the relationship between $X$ and $Y$ can be learned directly from observed data. Its performance is an empirical reference rather than a theoretical upper bound.

The benchmark also provides a useful statistical reference for distributional validity. Under squared-error loss, the population prediction function is the conditional expectation,

\begin{equation}
f^{*}(X)=E[Y\mid X],
\label{eq:conditional_mean}
\end{equation}

while the law of total variance gives

\begin{equation}
\operatorname{Var}(Y)
=
\operatorname{Var}\left(E[Y\mid X]\right)
+
E\left[\operatorname{Var}(Y\mid X)\right].
\label{eq:total_variance}
\end{equation}

Conditional-mean predictions can therefore be substantially less dispersed than observed outcomes whenever relevant individual variation remains unexplained by $X$. Distributional compression is consequently not unique to LLMs or evidence of failure by itself. Conversely, greater dispersion in an LLM-generated sample is informative only when it is accompanied by adequate calibration and distributional shape.

The supervised reference is also useful when interpreting substantive structure, since it shows which relationships with $X$ are recoverable through direct supervised learning. For comparability, the ML models and LLMs use the same substantive respondent-information base, although represented in tabular and textual form, respectively. Final ML predictions are generated out of sample, and several learning algorithms are used to avoid making the benchmark dependent on a single model class. The implementation of these requirements is described in the Supervised Machine-Learning Benchmarks subsection.

\subsection{Operational pipeline}
\label{subsec:validation_pipeline}

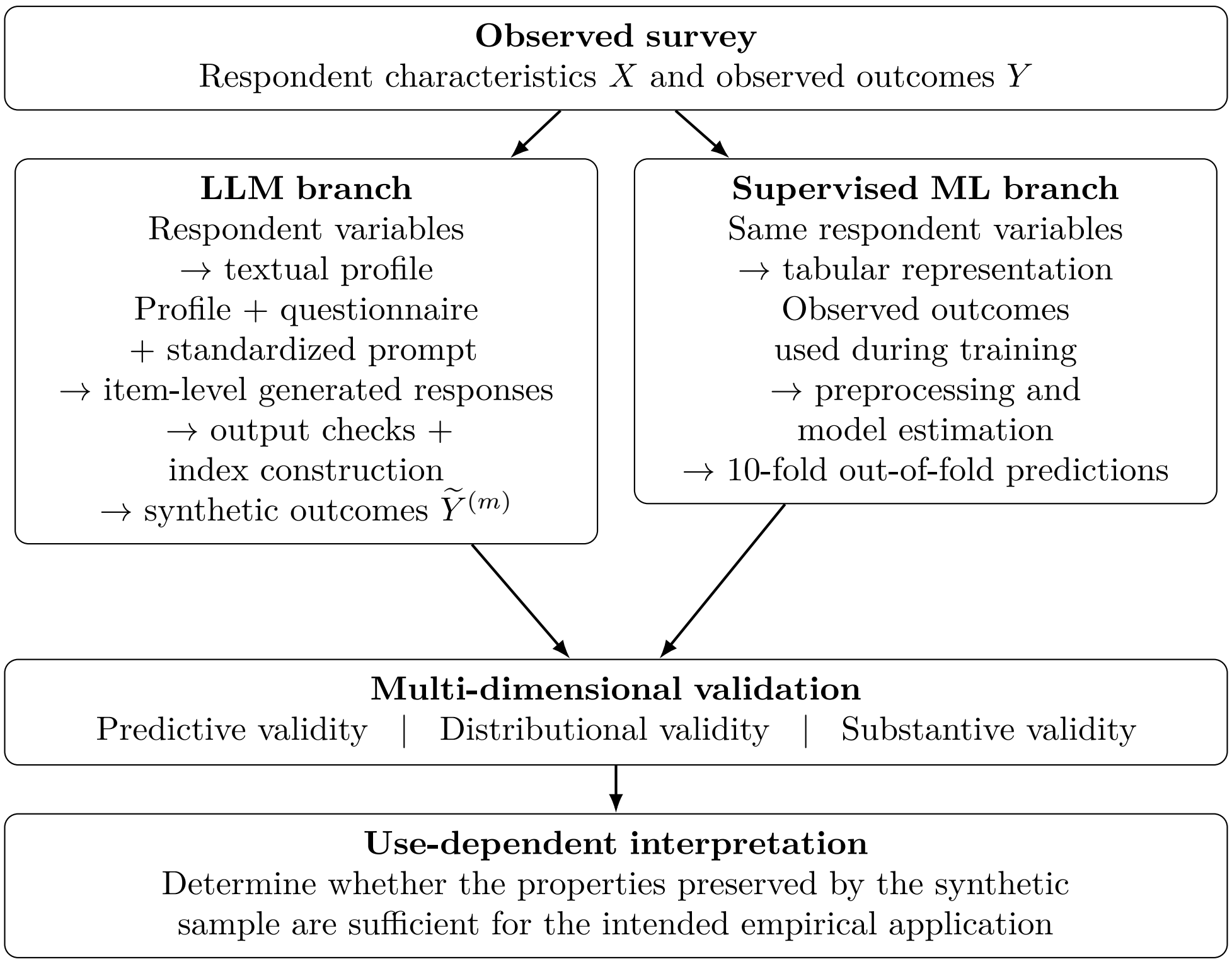
\begin{figure*}[t]
\centering

\begin{tikzpicture}[
    node distance=5mm,
    stage/.style={
        draw,
        rounded corners,
        align=center,
        inner sep=5pt,
        text width=0.78\textwidth,
        font=\small
    },
    branch/.style={
        draw,
        rounded corners,
        align=center,
        inner sep=5pt,
        text width=0.36\textwidth,
        font=\small
    },
    arrow/.style={
        -{Latex[length=2mm]},
        thick
    }
]

\node[stage] (input) {
\textbf{Observed survey}\\
Respondent characteristics $X$ and observed outcomes $Y$
};

\node[branch, below=of input, xshift=-3.25cm] (llm) {
\textbf{LLM branch}\\
Respondent variables $\rightarrow$ textual profile\\
Profile + questionnaire + standardized prompt\\
$\rightarrow$ item-level generated responses\\
$\rightarrow$ output checks + index construction\\
$\rightarrow$ synthetic outcomes $\widetilde{Y}^{(m)}$
};

\node[branch, below=of input, xshift=3.25cm] (ml) {
\textbf{Supervised ML branch}\\
Same respondent variables $\rightarrow$ tabular representation\\
Observed outcomes used during training\\
$\rightarrow$ preprocessing and model estimation\\
$\rightarrow$ 10-fold out-of-fold predictions
};

\node[stage, below=12mm of llm, xshift=3.25cm] (validation) {
\textbf{Multi-dimensional validation}\\
Predictive validity $\;|\;$ Distributional validity $\;|\;$ Substantive validity
};

\node[stage, below=of validation] (interpretation) {
\textbf{Use-dependent interpretation}\\
Determine whether the properties preserved by the synthetic sample are sufficient for the intended empirical application
};

\draw[arrow] (input) -- (llm);
\draw[arrow] (input) -- (ml);
\draw[arrow] (llm) -- (validation);
\draw[arrow] (ml) -- (validation);
\draw[arrow] (validation) -- (interpretation);

\end{tikzpicture}

\caption{Operational pipeline for validating LLM-generated survey samples}
\label{fig:validation_pipeline}
\end{figure*}

Figure~\ref{fig:validation_pipeline} summarizes the empirical implementation of the framework. The observed survey provides respondent characteristics $X$ and reference outcomes $Y$. The same substantive respondent information feeds two distinct branches. In the LLM branch, respondent variables are translated into textual profiles, combined with the questionnaire and a standardized prompt, and used to generate item-level responses. These outputs are checked and processed using the same coding and aggregation rules as the observed survey to construct synthetic outcomes. In the supervised branch, the same respondent information is represented in tabular form and used together with observed targets during training to produce out-of-sample predictions. Both sets of generated outcomes are then evaluated against the observed data along the three validation dimensions and interpreted according to the intended empirical use.

The pipeline makes explicit the different informational roles of the three empirical objects. Observed outcomes provide the reference against which validity is assessed; supervised predictions show what can be recovered when target outcomes are available for learning; and LLM-generated outcomes are produced without target-specific supervised training or calibration.

\section{Empirical Application}
\label{sec:data}

\subsection{Data, outcomes, and analytical samples}
\label{subsec:outcome_indices}

To illustrate this validation framework and show its usefulness for future research on silicon samples within the social sciences, we have carried out an empirical application on the topic of individuals' attitudes towards AI. The application uses Special Eurobarometer 554, \emph{Artificial Intelligence and the Future of Work}, fielded as part of Eurobarometer wave 101.4 between 25 April and 22 May 2024 across the 27 Member States of the European Union \citep{eurobarometer1014}. We restrict the analysis to respondents aged 18--65 and exclude respondents classified as retired, yielding 16,835 observations before item- and covariate-completeness requirements are imposed.

The main analysis considers three composite outcomes capturing different dimensions of attitudes toward artificial intelligence and digital technologies. In our harmonized item coding, the \emph{General} index combines five items (Q01--Q05) measuring broad evaluations of AI and recent digital technologies. The \emph{Effects} index combines five items (Q06--Q10) concerning their perceived consequences and usefulness in working life. The \emph{Use} index combines eight items (Q11--Q18) covering attitudes toward concrete workplace applications of AI, including recruitment, task allocation, personal-data processing, worker safety, monitoring, performance assessment, and automated dismissal.

All items use four-category response scales. Coding is harmonized so that higher values consistently indicate more favorable attitudes toward AI and digital technologies. Each index is computed as the unweighted mean of its constituent items, retaining the original 1--4 scale. The item grouping follows the conceptual organization of the questionnaire and is supported by correlation and principal-component diagnostics. Full item wording and index-construction diagnostics are reported in the online supplement.

The \emph{Effects} and \emph{Use} analyses use the common work-related complete-case sample defined during data preparation, yielding 12,843 respondents. \emph{General} is defined on the nested sample of 11,114 respondents with complete Q01--Q05 responses. Requiring complete information on the respondent characteristics used in the LLM profiles and supervised benchmarks then yields the final analytical samples of 12,583 respondents for \emph{Effects} and \emph{Use} and 10,935 for \emph{General}. All observed, LLM-generated, and ML-predicted outcomes are compared within these same outcome-specific samples. Further details on outcome construction and analytical samples are reported in Section B.2 of the online supplement.

\subsection{Respondent profiles and LLM generation}
\label{subsec:respondent_profiles}

For each respondent, we construct a structured profile from characteristics observed in the Eurobarometer data. The profiles contain socio-demographic and household information, including country, age, gender, community type, citizenship background, education, and household composition; labour-market characteristics, including occupation, employment status, sector, workplace size, and exposure to AI and digital technologies at work; and additional economic, behavioral, political, and institutional characteristics, including financial difficulties, life satisfaction, internet use, political interest, views of the European Union, and institutional trust.

The variables are supplied to the LLMs through interpretable textual descriptions. No observed response entering the \emph{General}, \emph{Effects}, or \emph{Use} indices, and no quantity constructed from those responses, is included in the profile. The supervised ML benchmarks use the same substantive respondent-information base in tabular form, as described in the Supervised Machine-Learning Benchmarks subsection. Full variable definitions and the respondent-profile template are provided in the online supplement.

Synthetic responses are generated using four off-the-shelf general-purpose LLMs from different providers: Claude Sonnet 4.6, GPT-4.1 mini, Gemini 3 Flash, and Grok 4.1 Fast Non-Reasoning. The models are used without task-specific fine-tuning or supervised calibration on the target Eurobarometer outcomes. The selection reflects a practically relevant research setting, using recent models that combined general-purpose capability with computational cost and latency compatible with repeated generation across thousands of respondent profiles.

The same standardized prompting strategy is used across providers. Following a neutral design \citep{abbink2006neutral}, each model receives the respondent profile together with the relevant questionnaire items and admissible response categories, without examples of preferred answers, information on observed response distributions, or other target-specific calibration. Keeping the substantive prompt structure common across providers is intended to prevent model comparisons from being confounded by provider-specific prompting choices. For each respondent, the complete profile and question set are submitted together in a single generation request, so that all item responses are produced under a common respondent context and form a single survey-response vector, mirroring the one-response-per-respondent structure of the observed survey.

The main analysis uses the temperature setting \(T=0.5\) for all four providers, a moderate setting intended to avoid both deterministic generation and excessive stochasticity \citep{sarstedt2024silicon}. Temperature is implemented on provider-specific scales and therefore the same value should not be interpreted as implying identical effective randomness across models. Our objective is to apply a common, moderate setting rather than to equate the stochastic properties of the different APIs. To examine within-model sensitivity to this choice, we also generated complete synthetic samples for GPT, Gemini, and Grok at \(T=0.2\) and \(T=1.0\). The alternative-temperature generations produced only limited differences in this application and did not materially alter the qualitative validation results. The \(T=0.5\) specification is therefore retained for the main analysis. Claude was not regenerated at the alternative temperatures because repeating the complete extraction involved substantially higher computational cost. Detailed temperature-robustness results are reported in the online supplement.

Models are instructed to return item-level numerical responses on the admissible 1--4 scales in structured form. Valid responses are not otherwise modified or calibrated after generation. The resulting item-level responses are processed using exactly the same grouping, coding direction, and aggregation rules applied to the observed survey data. Each respondent is therefore associated with an observed outcome \(Y_i\) and, for each LLM \(m\), a corresponding synthetic outcome \(\widetilde{Y}_i^{(m)}\) on the same analytical scale. This produces one paired synthetic sample for each LLM and outcome. Because the respondent characteristics are inherited from the observed survey, the validation exercise isolates the generation of outcomes conditional on the observed profiles rather than the generation of the population composition itself.

\subsection{Supervised Machine-Learning Benchmarks}
\label{subsec:ml}

We construct supervised benchmarks separately for \emph{General}, \emph{Effects}, and \emph{Use} using four learning approaches: Elastic Net, Random Forest, Histogram Gradient Boosting, and Super Learner. The models span regularized linear, tree-based, boosting, and ensemble methods \citep{zou2005elasticnet,breiman2001random,friedman2001gradient,wolpert1992stacked,vanderlaan2007super}, reducing dependence of the supervised reference on a single functional form or learning algorithm.

The supervised models use the same substantive respondent-information base represented in the LLM profiles, although encoded in tabular rather than textual form. Categorical predictors are one-hot encoded, continuous and ordered predictors enter numerically, and predictors are standardized where required by the estimation pipeline. No covariate imputation is performed: the benchmarks use the same complete-case respondent samples defined for the LLM comparison. Country differences are handled through fixed-effect residualization: country components are partialled out before model estimation and added back after prediction so that final predictions are expressed on the original 1--4 outcome scale. Full predictor definitions, encoding choices, and preprocessing procedures are reported in the online supplement.

All reported ML outcomes are 10-fold out-of-fold predictions. Each respondent is predicted by a model estimated without that respondent, avoiding the use of in-sample fitted values as the comparison with the LLM-generated outcomes. Transformations required for prediction are estimated on the corresponding training data and then applied to held-out observations. Hyperparameters are selected by cross-validation using mean squared error as the prediction criterion; the Super Learner combines Elastic Net, Random Forest, and Histogram Gradient Boosting through cross-validated stacking. Model-specific tuning procedures, parameter grids, and reproducibility details are provided in the online supplement.

The benchmark uses the same outcome-specific analytical samples as the LLM exercise: \(N=10{,}935\) for \emph{General} and \(N=12{,}583\) for \emph{Effects} and \emph{Use}. Observed outcomes, LLM-generated outcomes, and supervised predictions can therefore be compared for the same respondents and on the same analytical scale. As noted in the Supervised machine learning as a reference subsection, the supervised models observe target outcomes during training whereas the LLMs do not; this asymmetry defines the benchmark rather than a comparison between equivalent prediction procedures.

\section{Results}
\label{sec:results}

\subsection{Predictive Validity}
\label{subsec:predictive}

We begin with predictive validity, comparing observed and generated outcomes at the respondent level. Table~\ref{tab:predictive_validity} reports MSE, RMSE, and Pearson correlations for the four supervised benchmarks and the four LLM-generated samples. As these diagnostics were defined in the Predictive validity subsection of the Validation Framework, we focus here on their empirical interpretation.

\begin{table*}[t]
\small\sf\centering
\caption{Predictive validity: supervised ML benchmarks and LLM-generated outcomes}
\label{tab:predictive_validity}

\renewcommand{\arraystretch}{1.10}
\setlength{\tabcolsep}{5pt}

\begin{tabular}{llccc}
\toprule
\textbf{Outcome} & \textbf{Source} & \textbf{MSE} & \textbf{RMSE} & \textbf{Pearson corr.} \\
\midrule

General
& Elastic Net & 0.259 & 0.509 & 0.477 \\
& Random Forest & 0.246 & 0.496 & 0.517 \\
& HGB & 0.250 & 0.500 & 0.506 \\
& Super Learner & 0.248 & 0.498 & 0.510 \\
& Claude & 0.292 & 0.540 & 0.403 \\
& GPT & 0.364 & 0.603 & 0.376 \\
& Gemini & 0.335 & 0.578 & 0.390 \\
& Grok & 0.627 & 0.792 & 0.372 \\
\addlinespace

Effects
& Elastic Net & 0.285 & 0.533 & 0.459 \\
& Random Forest & 0.272 & 0.522 & 0.496 \\
& HGB & 0.276 & 0.525 & 0.485 \\
& Super Learner & 0.275 & 0.524 & 0.488 \\
& Claude & 0.332 & 0.576 & 0.354 \\
& GPT & 0.389 & 0.624 & 0.354 \\
& Gemini & 0.374 & 0.612 & 0.369 \\
& Grok & 0.635 & 0.797 & 0.348 \\
\addlinespace

Use
& Elastic Net & 0.313 & 0.559 & 0.518 \\
& Random Forest & 0.302 & 0.550 & 0.541 \\
& HGB & 0.307 & 0.554 & 0.531 \\
& Super Learner & 0.305 & 0.552 & 0.535 \\
& Claude & 0.437 & 0.661 & 0.362 \\
& GPT & 0.525 & 0.725 & 0.372 \\
& Gemini & 0.499 & 0.706 & 0.356 \\
& Grok & 0.895 & 0.946 & 0.288 \\
\bottomrule
\end{tabular}

\begin{minipage}{0.80\textwidth}
\footnotesize\rmfamily
\textit{Notes:} HGB denotes Histogram Gradient Boosting. ML values are based on out-of-fold predictions. LLM values refer to the synthetic indices generated at \(T=0.5\). All comparisons use the common outcome-specific analytical samples.
\end{minipage}

\end{table*}

The supervised models consistently provide the strongest respondent-level predictions. Random Forest records the lowest MSE for all three outcomes, although differences among the four ML benchmarks are small. The same pattern emerges from Pearson correlations, confirming that the supervised models recover more of the observed variation across respondents.

Among the LLMs, Claude achieves the lowest MSE for all three outcomes, while the ranking by correlation varies somewhat across models. Nevertheless, none of the LLM-generated samples reaches the predictive performance of the supervised benchmarks. The gap is particularly pronounced for \emph{Use}, which is also the most difficult outcome for the LLMs in absolute prediction error.

Predictive validity therefore provides a clear first result: conditioning general-purpose LLMs on rich respondent profiles recovers some systematic individual-level variation, but in this application it does not match the accuracy obtainable when the same respondent information is used in supervised models trained directly on the observed outcomes. Whether weaker individual prediction also implies poorer synthetic-sample quality on other dimensions is a separate question, to which we turn in the distributional analysis.

\subsection{Distributional Validity}
\label{subsec:distributional}

Distributional validity provides a different perspective from respondent-level prediction. Table~\ref{tab:distributional_validity} reports the six diagnostics defined in the Distributional validity subsection of the Validation Framework for the same four supervised benchmarks and four LLM-generated samples considered above.

\begin{table*}[t]
\small\sf\centering
\caption{Distributional validity: supervised ML benchmarks and LLM-generated outcomes}
\label{tab:distributional_validity}

\renewcommand{\arraystretch}{1.10}
\setlength{\tabcolsep}{5pt}

\begin{tabular}{llrrrrrr}
\toprule
\textbf{Outcome} &
\textbf{Source} &
\textbf{Bias} &
\textbf{SD ratio} &
\textbf{KS} &
\textbf{Wass.} &
\textbf{Mean q-err.} &
\textbf{Max q-err.} \\
\midrule

General
& Elastic Net     &  0.000 & 0.479 & 0.215 & 0.225 & 0.243 & 0.546 \\
& Random Forest   & -0.005 & 0.476 & 0.227 & 0.229 & 0.252 & 0.540 \\
& HGB             & -0.001 & 0.532 & 0.202 & 0.207 & 0.231 & 0.512 \\
& Super Learner   &  0.001 & 0.490 & 0.206 & 0.220 & 0.239 & 0.535 \\
& Claude          &  0.070 & 0.540 & 0.250 & 0.221 & 0.200 & 0.400 \\
& GPT             & -0.020 & 0.852 & 0.228 & 0.119 & 0.160 & 0.200 \\
& Gemini          & -0.009 & 0.778 & 0.255 & 0.171 & 0.200 & 0.400 \\
& Grok            & -0.179 & 1.329 & 0.270 & 0.250 & 0.240 & 0.400 \\
\addlinespace

Effects
& Elastic Net     &  0.000 & 0.465 & 0.238 & 0.251 & 0.264 & 0.457 \\
& Random Forest   & -0.004 & 0.454 & 0.255 & 0.260 & 0.282 & 0.464 \\
& HGB             &  0.000 & 0.480 & 0.240 & 0.249 & 0.275 & 0.455 \\
& Super Learner   &  0.002 & 0.464 & 0.236 & 0.251 & 0.269 & 0.457 \\
& Claude          & -0.065 & 0.544 & 0.258 & 0.268 & 0.240 & 0.400 \\
& GPT             &  0.068 & 0.793 & 0.158 & 0.113 & 0.120 & 0.400 \\
& Gemini          &  0.146 & 0.709 & 0.212 & 0.214 & 0.240 & 0.400 \\
& Grok            & -0.424 & 0.969 & 0.447 & 0.427 & 0.440 & 0.600 \\
\addlinespace

Use
& Elastic Net     &  0.000 & 0.518 & 0.196 & 0.259 & 0.229 & 0.476 \\
& Random Forest   &  0.000 & 0.548 & 0.173 & 0.247 & 0.220 & 0.437 \\
& HGB             &  0.001 & 0.526 & 0.185 & 0.257 & 0.230 & 0.456 \\
& Super Learner   &  0.002 & 0.524 & 0.191 & 0.257 & 0.227 & 0.470 \\
& Claude          & -0.255 & 0.378 & 0.401 & 0.408 & 0.350 & 0.625 \\
& GPT             & -0.272 & 0.813 & 0.376 & 0.301 & 0.300 & 0.750 \\
& Gemini          & -0.331 & 0.550 & 0.362 & 0.383 & 0.375 & 0.625 \\
& Grok            & -0.686 & 0.562 & 0.631 & 0.696 & 0.725 & 1.000 \\
\bottomrule
\end{tabular}

\begin{minipage}{0.80\textwidth}
\footnotesize\rmfamily
\textit{Notes:} HGB denotes Histogram Gradient Boosting. Bias is defined as the predicted or synthetic mean minus the observed mean. The SD ratio is the predicted or synthetic standard deviation divided by the observed standard deviation; values below one indicate compression and values above one excess dispersion. KS denotes the Kolmogorov--Smirnov distance and Wass.\ the Wasserstein distance. Mean q-err.\ is the mean absolute quantile error across the 10th, 25th, 50th, 75th, and 90th percentiles, while Max q-err.\ is the largest absolute error across the same percentiles. Lower values indicate closer fit for the distance and quantile-error measures. ML values are based on out-of-fold predictions. LLM values refer to \(T=0.5\).
\end{minipage}

\end{table*}

The supervised benchmarks reproduce the observed means almost exactly, but their predicted distributions are substantially compressed. Across the three outcomes, SD ratios remain close to one half of the observed dispersion. This pattern is consistent with the conditional-mean interpretation of supervised prediction discussed in the Supervised machine learning as a reference subsection: strong prediction of the systematic component of an outcome does not imply recovery of its full marginal heterogeneity.

The LLM-generated samples display considerably greater variation across models. For \emph{General} and \emph{Effects}, GPT and Gemini preserve substantially more dispersion than the supervised benchmarks, whereas Claude remains much closer to the compressed dispersion of the supervised predictions. The comparison with the predictive results is particularly informative in this respect. Claude has the lowest respondent-level prediction error among the LLMs, but GPT provides the closest fit on several distributional diagnostics. For \emph{General}, GPT records the lowest Wasserstein distance and the smallest mean and maximum quantile errors. For \emph{Effects}, it records the lowest Kolmogorov--Smirnov and Wasserstein distances and the smallest mean quantile error.

Greater dispersion does not, however, imply greater distributional validity by itself. Grok provides the clearest example for \emph{General}: its synthetic distribution is more dispersed than the observed distribution, with an SD ratio of 1.329, but this is accompanied by a sizeable negative mean bias and comparatively large distributional distances. Additional synthetic variation can therefore reflect miscalibration rather than meaningful heterogeneity.

The pattern is less favorable for \emph{Use}. All four LLMs generate substantially negative mean biases, and their distributional-distance measures are generally larger than those of the supervised benchmarks. GPT retains considerably more dispersion than the other models, but this does not eliminate its level and shape discrepancies. The difficulties already identified in respondent-level prediction therefore extend to the marginal distribution of this outcome.

Overall, the distributional results reveal a clear trade-off that would be missed by predictive diagnostics alone. Supervised ML provides very accurate mean calibration but systematically compresses the observed distribution. Some LLM-generated samples preserve more of the marginal dispersion and, for \emph{General} and \emph{Effects}, reproduce selected aspects of distributional shape more closely despite weaker individual-level prediction. At the same time, excess variation and poor calibration can coexist. Distributional validity must therefore be assessed jointly through location, dispersion, and shape rather than inferred from either predictive accuracy or variance alone. Corresponding density plots are reported in the online supplement.

\subsection{Substantive Validity}
\label{subsec:substantive}

Substantive validity asks whether generated outcomes preserve empirical relationships that would matter in downstream social-science analysis. Given the importance of empirical relationships among variables in social-science research, this dimension is fundamental, as it investigates whether and the extent to which synthetic data can be used to analyze and reproduce such relationships. We examine this property through three complementary exercises: heterogeneity patterns across age and education groups, cross-country variation, and similarity in predictor rankings.

\subsection*{Patterns across age and education groups}
\label{subsec:substantive_groups}

We first examine whether observed differences across standard demographic groups are preserved in the synthetic outcomes. For each outcome, we compare group means across age and education categories using three diagnostics: the correlation between observed and synthetic group means, their mean absolute error (MAE), and the ratio between the standard deviation of synthetic and observed group means (BG-SD ratio). The first diagnostic captures whether the ordering of groups is preserved, the second captures calibration in levels, and the third captures whether between-group heterogeneity is compressed or exaggerated. Because these correlations are computed across a small number of group means, we interpret them jointly with MAE and BG-SD rather than as standalone measures of substantive correspondence.

To keep the demographic comparison compact, we use Random Forest as the supervised reference in this exercise. As shown in the Predictive Validity subsection, the four supervised benchmarks yield broadly similar predictive performance, but Random Forest provides the strongest overall benchmark in this application, recording both the lowest MSE and the highest respondent-level correlation for each of the three outcomes.

\begin{table*}[t]
\small\sf\centering
\caption{Substantive validity across age and education groups}
\label{tab:group_validity}

\renewcommand{\arraystretch}{1.10}
\setlength{\tabcolsep}{5pt}

\begin{tabular}{llccc|ccc}
\toprule
&
&
\multicolumn{3}{c}{\textbf{Age groups}} &
\multicolumn{3}{c}{\textbf{Education groups}} \\
\cmidrule(lr){3-5}
\cmidrule(lr){6-8}
\textbf{Outcome} &
\textbf{Source} &
\textbf{Corr.} &
\textbf{MAE} &
\textbf{BG-SD} &
\textbf{Corr.} &
\textbf{MAE} &
\textbf{BG-SD} \\
\midrule

General
& Random Forest & 0.998 & 0.010 & 0.826 & 1.000 & 0.012 & 0.900 \\
& Claude        & 0.992 & 0.072 & 1.285 & 0.987 & 0.077 & 1.056 \\
& GPT           & 0.988 & 0.027 & 1.408 & 0.967 & 0.096 & 1.799 \\
& Gemini        & 0.991 & 0.032 & 1.558 & 0.995 & 0.056 & 1.485 \\
& Grok          & 0.995 & 0.168 & 2.079 & 0.982 & 0.290 & 3.363 \\
\addlinespace

Effects
& Random Forest & 0.998 & 0.011 & 0.798 & 0.999 & 0.012 & 0.883 \\
& Claude        & 0.991 & 0.085 & 2.146 & 0.987 & 0.057 & 1.116 \\
& GPT           & 0.985 & 0.071 & 1.414 & 0.982 & 0.105 & 1.895 \\
& Gemini        & 0.992 & 0.155 & 1.896 & 0.998 & 0.152 & 1.772 \\
& Grok          & 0.982 & 0.419 & 1.654 & 0.996 & 0.417 & 2.625 \\
\addlinespace

Use
& Random Forest & 0.999 & 0.009 & 0.891 & 0.998 & 0.006 & 0.916 \\
& Claude        & 0.985 & 0.262 & 0.606 & 0.953 & 0.245 & 1.331 \\
& GPT           & 0.881 & 0.283 & 0.519 & 0.911 & 0.251 & 2.415 \\
& Gemini        & 0.978 & 0.339 & 0.569 & 0.943 & 0.316 & 1.996 \\
& Grok          & 0.825 & 0.698 & 0.518 & 0.928 & 0.669 & 2.824 \\
\bottomrule
\end{tabular}

\begin{minipage}{0.80\textwidth}
\footnotesize\rmfamily
\textit{Notes:} Corr.\ is the correlation between observed and predicted or synthetic group means. MAE is the mean absolute error of group means. BG-SD is the standard deviation of predicted or synthetic group means divided by the standard deviation of observed group means. Values below one indicate compression of between-group heterogeneity; values above one indicate exaggeration. Random Forest values are based on out-of-fold predictions. LLM values refer to \(T=0.5\).
\end{minipage}

\end{table*}

The Random Forest benchmark reproduces demographic group means very closely. Correlations with the observed group means range from 0.998 to 1.000 and MAE values from 0.006 to 0.012. Its BG-SD ratios are consistently below one, however, indicating mild compression of the observed differences across both age and education groups.

The LLM-generated outcomes also recover a substantial amount of demographic structure. For \emph{General} and \emph{Effects}, correlations between observed and synthetic group means are generally above 0.98 for both age and education. This means that the models largely preserve the ordering of demographic groups without task-specific training or calibration on the target outcomes. Their calibration in levels is weaker than that of Random Forest, however, and the magnitude of between-group differences is not always reproduced accurately.

The distinction is particularly visible in the BG-SD ratios. For \emph{General} and \emph{Effects}, the LLMs generally exaggerate differences across age groups. Differences across education groups are also typically amplified, with Claude providing the main exception. For \emph{Use}, the pattern differs: all four LLMs compress age-related heterogeneity while amplifying differences across education groups. Overall, Claude is generally closest to the observed between-group dispersion, although its performance varies across outcomes and grouping variables.

\subsection*{Cross-country patterns}
\label{subsec:substantive_country}

We next examine whether the LLM-generated outcomes preserve cross-country variation in the observed survey. For each of the 27 EU Member States, we compute the observed and synthetic mean of each outcome and then calculate the correlation between observed and LLM-generated country means.

We report this diagnostic only for the LLM-generated outcomes. In the supervised ML benchmark, country effects are explicitly partialled out during model estimation and the estimated country component is subsequently added back to the out-of-fold predictions. Cross-country differences are therefore incorporated directly into the construction of the supervised predictions, making their country-level correspondence with the observed outcomes partly mechanical. The same diagnostic would consequently not provide a comparable validation test for ML and LLM outputs. By contrast, the LLMs receive country only as one element of the respondent profile and are not calibrated to reproduce the observed country means.

\begin{table*}[t]
\small\sf\centering
\caption{Correspondence between observed and LLM-generated country means}
\label{tab:country_validity}

\renewcommand{\arraystretch}{1.10}
\setlength{\tabcolsep}{5pt}

\begin{tabular}{lccc}
\toprule
\textbf{LLM} &
\textbf{General} &
\textbf{Effects} &
\textbf{Use} \\
\midrule
Claude & 0.702 & 0.630 & 0.653 \\
GPT    & 0.480 & 0.401 & 0.185 \\
Gemini & 0.727 & 0.611 & 0.599 \\
Grok   & 0.580 & 0.504 & 0.193 \\
\bottomrule
\end{tabular}

\begin{minipage}{0.80\textwidth}
\footnotesize\rmfamily
\textit{Notes:} Entries are Pearson correlations between observed and LLM-generated mean outcomes across the 27 EU Member States. LLM values refer to synthetic indices generated at \(T=0.5\) on the corresponding analytical samples.
\end{minipage}

\end{table*}

The country-level results show that some LLM-generated samples preserve a meaningful part of the geographic variation observed in the survey. Gemini performs best for \emph{General}, with a correlation of 0.727, followed closely by Claude at 0.702. Claude provides the strongest correspondence for \emph{Effects} and \emph{Use}, with correlations of 0.630 and 0.653, respectively, while Gemini also retains substantial cross-country structure for both outcomes.

The result is particularly informative because the LLMs are not calibrated to the observed country means. At the same time, country-level correspondence should not be confused with overall calibration. For example, Claude recovers a substantial part of the cross-country variation in \emph{Use} despite the negative mean bias documented in the Distributional Validity subsection. The country-level diagnostic therefore provides evidence of preserved geographic structure, but not of accurate outcome levels.

GPT and Grok show substantially weaker correspondence for \emph{Use}, with country-level correlations of 0.185 and 0.193. The preservation of geographic structure is therefore strongly model- and outcome-dependent, reinforcing the broader conclusion that substantive validity cannot be inferred from the general use of an LLM alone.

\subsection*{Predictor-structure similarity}
\label{subsec:substantive_predictors}

The final substantive-validity exercise asks whether the LLM-generated outcomes preserve the predictor structure that emerges from the observed survey data. The intuition is to ask whether a researcher analysing the synthetic outcomes, rather than the observed survey outcomes, would identify broadly similar respondent characteristics as empirically important.

For this exercise, Elastic Net is used not as an alternative supervised benchmark, but as a common diagnostic model applied ex post to the observed and already-generated synthetic outcomes. The LLMs are not queried or regenerated during this exercise: for each model, the synthetic outcome vector is held fixed. We first estimate the full Elastic Net specification using the observed outcome and then remove one predictor at a time, recording the resulting increase in out-of-fold MSE. Larger increases indicate a stronger predictive contribution of the omitted predictor. We then repeat the same ablation procedure using each fixed LLM-generated outcome as the dependent variable, while holding the covariates, analytical sample, and procedure constant. The resulting predictor-importance rankings can therefore be compared to assess whether observed and synthetic outcomes imply similar predictor structures. These rankings are predictive rather than causal.

We compare each LLM-based ranking with the ranking obtained from the observed outcome using Spearman rank correlations and overlap among the five and ten highest-ranked predictors. Table~\ref{tab:predictor_similarity} reports the results.

\begin{table*}[t]
\small\sf\centering
\caption{Similarity between observed and LLM-based predictor rankings}
\label{tab:predictor_similarity}

\renewcommand{\arraystretch}{1.10}
\setlength{\tabcolsep}{5pt}

\begin{tabular}{llccc}
\toprule
\textbf{Outcome} &
\textbf{LLM} &
\textbf{Spearman corr.} &
\textbf{Top-5 overlap} &
\textbf{Top-10 overlap} \\
\midrule

General
& Claude & 0.787 & 4/5 & 9/10 \\
& GPT    & 0.711 & 3/5 & 9/10 \\
& Gemini & 0.756 & 4/5 & 9/10 \\
& Grok   & 0.589 & 3/5 & 8/10 \\
\addlinespace

Effects
& Claude & 0.858 & 4/5 & 8/10 \\
& GPT    & 0.664 & 3/5 & 8/10 \\
& Gemini & 0.800 & 4/5 & 8/10 \\
& Grok   & 0.713 & 3/5 & 9/10 \\
\addlinespace

Use
& Claude & 0.717 & 4/5 & 8/10 \\
& GPT    & 0.579 & 3/5 & 7/10 \\
& Gemini & 0.564 & 4/5 & 7/10 \\
& Grok   & 0.585 & 3/5 & 8/10 \\
\bottomrule
\end{tabular}

\begin{minipage}{0.80\textwidth}
\footnotesize\rmfamily
\textit{Notes:} The reference ranking is obtained by applying Elastic Net predictor-ablation analysis to the observed outcome. The same procedure is then applied to each LLM-generated outcome using the same covariates and analytical sample. Spearman correlations measure similarity across the complete rankings. Top-5 and Top-10 overlap report the number of predictors shared with the corresponding observed-data ranking. LLM values refer to \(T=0.5\).
\end{minipage}

\end{table*}

The LLM-generated outcomes preserve a substantial part of the predictor structure identified in the observed data. Spearman correlations are moderate to high across models and outcomes, and every synthetic sample recovers at least seven of the ten highest-ranked predictors. Claude provides the closest overall correspondence for all three outcomes, with rank correlations of 0.787 for \emph{General}, 0.858 for \emph{Effects}, and 0.717 for \emph{Use}. Gemini also performs strongly for \emph{General} and \emph{Effects}.

The detailed rankings reveal substantive similarities as well as meaningful discrepancies. Institutional trust remains prominent in the synthetic outcomes, particularly trust in European Union institutions, which is also a leading predictor in the observed-data analysis. The correspondence is less uniform for direct exposure to AI and digital technologies at work. For example, workplace AI exposure is the leading predictor of \emph{Use} in the observed-data ranking, whereas it ranks fourth in the Claude-based ranking. The complete predictor-level rankings are reported in the online supplement.

Taken together, the three substantive-validity exercises show that weaker respondent-level prediction does not imply that LLM-generated outcomes miss empirically meaningful structure. Some synthetic samples recover demographic ordering, cross-country variation, and a substantial part of the predictor structure observed in the survey. At the same time, the magnitude and content of these relationships are not reproduced uniformly across models or outcomes. Substantive agreement should therefore be treated as a distinct validation object rather than inferred from predictive or distributional performance alone.

\section{Discussion}
\label{sec:discussion}

The results of this empirical application support the central motivation for a use-dependent multi-dimensional validation framework: the value of an LLM-generated synthetic sample cannot be characterized adequately by a single measure of agreement with observed data. Predictive, distributional, and substantive validity capture genuinely different properties, and the same synthetic sample can perform differently across them. In our application, this divergence across validation dimensions shows concretely why the suitability of a synthetic sample depends on the empirical property required by its intended use. Evaluating synthetic respondents only through individual prediction error would overlook forms of distributional and substantive structure that may be relevant for social-science applications.

The comparison with supervised machine learning provides a demanding but informative reference. The ML models are trained directly on the observed target outcomes and are explicitly optimized for prediction, whereas the LLMs are used off the shelf, without task-specific training, fine-tuning, or supervised calibration on those outcomes. Their weaker respondent-level accuracy is therefore expected. More importantly, the positive respondent-level correlations obtained by the LLMs show that general-purpose pre-trained models can recover systematic variation from respondent profiles and questionnaire information under a substantially less favorable learning setting. The ML benchmark consequently helps establish how much additional accuracy is obtained when the relationship between respondent characteristics and the observed target can be learned directly from the data.

The distributional results reveal a complementary trade-off. Supervised ML reproduces observed means very closely but generates substantially compressed predictions, consistent with its orientation toward conditional-mean prediction. Some LLM-generated samples preserve considerably more of the observed marginal dispersion and, for \emph{General} and \emph{Effects}, reproduce selected aspects of distributional shape more closely. This does not mean that greater dispersion is intrinsically preferable: additional synthetic variation can also coexist with bias or poor calibration. It does show, however, that LLM-generated outcomes are not simply less accurate versions of supervised predictions. They constitute a different empirical object, and their potential value cannot be assessed from predictive accuracy alone.

The most encouraging evidence concerns substantive validity. For the broader \emph{General} and \emph{Effects} attitudes, several LLM-generated samples recover much of the observed ordering across age and education groups, retain meaningful cross-country variation, and reproduce a substantial part of the predictor structure identified in the observed survey data. This is a notable result because these relationships emerge without task-specific training or calibration on the target outcomes. The generated responses are therefore not merely plausible in isolation: some of them preserve non-trivial demographic, geographic, and predictor-level structure present in the observed data.

The detailed substantive results also provide information about where this correspondence is stronger and where it becomes less reliable. Institutional trust remains prominent in the synthetic predictor rankings, while direct exposure to AI and digital technologies at work is reproduced less consistently relative to the observed-outcome analysis. Performance also varies across outcomes, with \emph{Use} proving substantially more difficult than the two broader attitude indices. This within-domain heterogeneity is itself informative: evidence that an LLM reproduces one construct should not be transferred automatically to another construct generated from the same respondent profiles.

Taken together, the evidence supports a cautiously optimistic interpretation. Off-the-shelf LLMs do not match supervised ML when the objective is calibrated respondent-level prediction, but some already recover meaningful empirical structure across several levels of analysis without task-specific training, fine-tuning, or supervised calibration on the target survey outcomes. Their usefulness therefore lies not only in generating plausible answers, but also in their capacity to reproduce selected features of observed social-scientific data from respondent profiles and questionnaire information.

This makes LLM-generated samples potentially useful as validation-dependent complements to observed data. Possible applications include exploratory analysis, questionnaire development, scenario simulation, hypothesis formation, and preliminary response modelling when target outcomes are unavailable or too limited to support supervised learning. More demanding applications, such as respondent-level replacement, reproduction of population heterogeneity, or substantive empirical inference, require correspondingly stronger evidence on the validation dimensions relevant to that use. The purpose of the framework is precisely to make these requirements explicit rather than to classify LLM-generated samples as either generally valid or generally invalid.

The empirical evidence nevertheless has clear scope conditions. The analysis concerns one large multi-country survey and one substantive domain, using four specific LLMs and a common profile-based prompting strategy. Model versions, prompt design, response formats, or generation settings may affect the resulting synthetic samples. The framework is intended to be portable across these choices, whereas the relative performance of the individual models reported here should be treated as application-specific. The exercise also conditions on respondent profiles taken from the observed Eurobarometer sample. It therefore evaluates the ability of LLMs to generate outcomes within an observed respondent-profile space, not their ability to generate the composition of a representative population from scratch. Applications in which both respondent characteristics and survey outcomes are synthetic would require additional validation of population composition and representativeness. Likewise, the predictor-ranking exercise is predictive rather than causal: similarity in predictor importance indicates preservation of empirical predictor structure, not preservation of causal effects.

A further extension concerns the use of synthetic observations to augment observed samples. Recent work has begun to develop formal strategies for combining human observations with imperfect LLM-generated predictions. \citet{broska2025mixed}, for example, propose a Mixed Subjects Design in which LLM predictions are treated as potentially informative observations while human subjects remain the empirical gold standard, allowing prediction-powered inference to exploit synthetic information without assuming that generated and observed responses are interchangeable. Our analysis addresses a complementary prior question: it evaluates which properties of the synthetic outcomes themselves are preserved, rather than whether combining synthetic and observed observations improves estimation for a particular empirical task. Whether such augmentation is beneficial will depend on the intended analysis, the amount and composition of observed data, and the errors introduced by the synthetic observations. The validation framework can therefore inform such downstream designs by identifying which properties of a synthetic sample should first be established for its intended use.

Overall, the results suggest that LLM-generated survey samples should neither be accepted because their answers appear plausible nor rejected simply because they underperform supervised prediction. The substantive evidence shows that current off-the-shelf models can already recover a non-trivial amount of empirical structure from respondent profiles and questionnaire information, without task-specific training or calibration on the target survey outcomes. The appropriate question is therefore which properties they preserve, for which outcomes, and whether those properties are sufficient for the intended empirical use.

\section{Conclusion}
\label{sec:conclusion}

This paper has proposed a use-dependent, multi-dimensional framework for evaluating LLM-generated synthetic survey samples in social-science applications. The framework distinguishes predictive, distributional, and substantive validity and links each validation object to the empirical purposes for which synthetic survey data may be used. Rather than seeking a single measure of synthetic-sample quality, it provides a common structure for assessing separately the properties that are relevant to a given intended use.

Applying the framework to a large multi-country Eurobarometer survey reveals precisely this pattern. Supervised machine-learning models provide stronger respondent-level prediction and close mean calibration, but substantially compress the observed outcome distributions. The LLM-generated samples are more heterogeneous: although they predict individual responses less accurately, some preserve more distributional variation and reproduce meaningful demographic, geographic, and predictor-level structure.

The substantive results are particularly encouraging. Several off-the-shelf LLMs recover substantial features of the observed social-scientific structure without task-specific training, fine-tuning, or supervised calibration on the target survey outcomes. At the same time, performance differs across models and constructs, showing that such correspondence cannot be assumed from the use of an LLM alone. Synthetic samples should therefore be evaluated for the specific outcome and empirical purpose for which they are intended.

The broader implication is that LLM-generated survey samples should be viewed neither as automatic substitutes for observed data nor simply as inferior versions of supervised predictions. They represent a distinct empirical resource whose usefulness depends on whether they preserve the properties required by the intended empirical application. The proposed use-dependent, multi-dimensional framework provides a reusable basis for identifying and evaluating those properties across models, outcomes, and applications, and for determining when synthetic respondents can serve as informative complements to conventional survey data.

\section*{Notes}
\noindent\textbf{Data source.} The source survey data are Eurobarometer 101.4 (ZA8844, version 1.0.0), available from the GESIS Data Archive at \href{https://doi.org/10.4232/1.14471}{doi:10.4232/1.14471}.

\noindent\textbf{Use of generative AI.} During the preparation of this manuscript, the authors used ChatGPT (OpenAI) to assist with language editing, drafting, and revision. The authors reviewed and edited all AI-assisted content and take full responsibility for the content of the publication. The use of LLMs to generate synthetic survey responses is described in the Methods and Supplementary Appendix.

\bibliographystyle{apalike}
\bibliography{bibl}

\clearpage
\appendix
\setcounter{table}{0}
\setcounter{figure}{0}
\renewcommand{\thetable}{A\arabic{table}}
\renewcommand{\thefigure}{A\arabic{figure}}
\renewcommand{\theHtable}{appendix.\thesection.\arabic{table}}
\renewcommand{\theHfigure}{appendix.\thesection.\arabic{figure}}
\section*{Supplementary Appendix}
This appendix provides additional literature documentation, outcome-construction details, robustness analyses, implementation information, and supplementary results supporting the empirical application in the main paper.
\section{Validation strategies in the related literature}
\label{app:validation_literature}

Table~\ref{tab:validation_literature} summarizes the main evaluation strategies adopted in a selected set of studies that are particularly relevant to the validation of LLM-generated respondents and survey-style social data. The table is not intended as an exhaustive review of the broader literature on LLM evaluation. Rather, it documents the heterogeneity of the empirical objects that closely related studies seek to reproduce or validate and clarifies how recent frameworks organize these objects. The final column provides an approximate mapping onto the dimensions proposed in the present paper; this mapping is our interpretation rather than terminology necessarily used by the original authors.

\begin{landscape}

\scriptsize
\setlength{\tabcolsep}{4pt}
\renewcommand{\arraystretch}{1.25}

\begin{longtable}{
    p{3.0cm}
    p{4.1cm}
    p{8.0cm}
    p{7.0cm}
}

\caption{Selected validation strategies for LLM-generated respondents and social data}
\label{tab:validation_literature}\\

\toprule
\textbf{Study} &
\textbf{Empirical setting} &
\textbf{Main validation or evaluation strategy} &
\textbf{Approximate relation to our framework} \\
\midrule
\endfirsthead

\multicolumn{4}{l}{\textit{Table \thetable\ continued}}\\
\toprule
\textbf{Study} &
\textbf{Empirical setting} &
\textbf{Main validation or evaluation strategy} &
\textbf{Approximate relation to our framework} \\
\midrule
\endhead

\midrule
\multicolumn{4}{r}{\textit{Continued on next page}}\\
\endfoot

\bottomrule
\endlastfoot

\citet{argyle2023outofone}
&
U.S. survey samples; GPT-3 conditioned on socio-demographic backstories
&
Develop \emph{algorithmic fidelity} through four criteria: a Social Science Turing Test, Backward Continuity, Forward Continuity, and Pattern Correspondence. The empirical exercises compare generated and human responses at both individual and subgroup levels and examine whether generated responses preserve relationships between demographic characteristics, attitudes, and behavior. The authors do not propose a universal numerical metric or threshold for satisfying the four criteria.
&
Spans several validation objects. Pattern Correspondence is particularly close to \emph{substantive validity}, while individual-level and subgroup comparisons overlap with predictive and distributional aspects of our framework. The approach is multidimensional but is not organized around an explicit mapping between validation dimensions and intended empirical uses.
\\

\addlinespace

\citet{aher2023using}
&
Replications of established economic, psycholinguistic, and social-psychology experiments
&
Introduce ``Turing Experiments,'' in which samples of LLM-simulated participants are evaluated according to whether they reproduce established empirical findings from human-subject research. Applications include the Ultimatum Game, Garden Path Sentences, the Milgram experiment, and Wisdom of Crowds. The comparison therefore focuses primarily on replication of previously documented behavioral patterns and on identifying systematic simulation distortions.
&
Closest to a downstream or \emph{substantive-validity} criterion: the central question is whether analyses based on simulated participants reproduce established human findings rather than whether each simulated individual matches a corresponding observed respondent.
\\

\addlinespace

\citet{bisbee2024synthetic}
&
Public-opinion feeling thermometers; synthetic personas benchmarked against ANES respondents
&
Compare synthetic and observed average responses, response variation, and conditional relationships estimated through regressions. They additionally examine sensitivity to small changes in prompt wording and changes in outputs generated with the same prompt over time. Their results illustrate that relatively accurate aggregate means can coexist with lower synthetic variance and substantially different regression coefficients.
&
Directly spans \emph{distributional validity} and \emph{substantive validity}, with additional reproducibility diagnostics. It provides a particularly clear example of why agreement in central tendency does not establish preservation of the relationships required for substantive analysis.
\\

\addlinespace

\citet{dominguez2024questioning}
&
American Community Survey questions answered by multiple LLMs and compared with human reference populations
&
Evaluate the distributions of model survey responses and their apparent alignment with demographic subgroups while systematically varying answer ordering and labeling. They examine entropy and other statistical signals in human and model-generated data and use a discriminator exercise to assess how readily synthetic observations can be distinguished from census observations. The analysis shows that apparent population alignment can be strongly affected by ordering and labeling biases.
&
Primarily \emph{distributional validity}, complemented by robustness and generation-procedure diagnostics. The study demonstrates that distributional similarity itself can be confounded by systematic properties of the prompting procedure.
\\

\addlinespace

\citet{boelaert2025machinebias}
&
Opinion-poll questions answered by generative language models and compared with human survey responses
&
Examine whether LLMs can emulate human respondents across a range of opinion and attitudinal questions. The analysis documents systematic discrepancies in response levels and substantially reduced within-topic variation, while also showing that the direction and magnitude of model bias vary across topics. The authors characterize this combination of systematic bias, low variance, and unstable topic-specific distortion as ``machine bias.''
&
Primarily \emph{distributional validity}, with implications for the use of synthetic respondents in substantive survey research. The study provides direct evidence that apparently plausible model responses may simultaneously distort both outcome levels and heterogeneity, reinforcing the need to evaluate location and dispersion separately rather than treating aggregate plausibility as sufficient validation.
\\

\addlinespace

\citet{lee2024climate}
&
U.S. public-opinion surveys on global warming; GPT models conditioned on demographic characteristics and, alternatively, additional issue-relevant covariates
&
Assess algorithmic fidelity by comparing LLM-generated and observed responses across several public-opinion questions. The study evaluates respondent-level prediction using accuracy and F1-based measures, compares generated and observed response distributions, and examines bias across population groups. A central comparison varies the conditioning information supplied to the models and shows that adding issue-relevant covariates can substantially improve fidelity for some outcomes.
&
Spans \emph{predictive validity} and \emph{distributional validity}, with subgroup-bias analyses related to \emph{substantive validity}. It also demonstrates that validation results can depend strongly on the information included in respondent profiles, reinforcing the need to evaluate a specific generation procedure rather than an LLM in the abstract.
\\

\addlinespace

\citet{yang2024large}
&
Social prediction tasks based on survey features, including voting and social-profile prediction
&
Assess respondent-level predictive performance under different information settings. The analysis uses Accuracy and Cohen's Kappa in the voting exercise and AUC in the broader social-profile prediction tasks. A central validation exercise removes ``shortcut'' input features that are closely related to the target and examines the resulting change in performance. LLM predictions are also compared with supervised Random Forest benchmarks.
&
Primarily \emph{predictive validity}, combined with an important information-set robustness test. The focus is on whether apparent respondent-level predictive success survives a more realistic restriction of the information provided to the model.
\\

\addlinespace

\citet{miranda2025simulating}
&
2020 ANES public-opinion questions; LLM simulations compared with Random Forest predictions
&
Explicitly separate respondent-level predictive performance from aggregate distributional performance. Individual-level evaluation uses weighted F1-score and Cram\'er's V, while Jensen--Shannon distance measures similarity between generated and observed opinion distributions. Performance is also compared with a supervised Random Forest under matched information conditions.
&
Directly corresponds to the distinction between \emph{predictive validity} and \emph{distributional validity}. The study is one of the closest antecedents to our separation of these two validation objects, but it does not formulate preservation of downstream analysis-relevant relationships as a third general validation dimension or link the dimensions systematically to intended uses.
\\

\addlinespace

\citet{sinacola2025llms}
&
World Values Survey; virtual populations generated from demographic profiles and compared with Random Forest
&
Evaluate predictive accuracy on survey questions, compare LLM performance with Random Forest models trained on different amounts of observed data, and examine prediction accuracy across geographic and religious groups. They also study sensitivity to prompt design and temperature and compare censored and uncensored model variants.
&
Primarily \emph{predictive validity}, with subgroup diagnostics related to \emph{substantive validity} and additional robustness checks concerning the generation procedure. The evaluation illustrates how aggregate predictive performance can conceal substantial heterogeneity across population groups.
\\

\addlinespace

\citet{lin2026psychological}
&
Methodological framework for using LLMs as psychological and behavioral simulators across individual, persona-based, and more complex simulation settings
&
Develops a validation framework whose first stage depends on the availability of human benchmark data. \emph{Direct validation} applies when comparable human data exist and includes task equivalence, response distributions, and process-level features; \emph{indirect validation} uses constituent processes, convergent evidence, and stress tests when only partial human evidence is available; and \emph{generative validation} relies on theoretical coherence, emergent properties, and expert assessment when corresponding human data cannot feasibly be obtained. The framework additionally emphasizes implementation confounds, interpretation of human--LLM discrepancies, and transparent reporting.
&
The framework addresses a broader simulation-validity problem and is organized primarily by the availability of human validation evidence rather than by the empirical use of synthetic survey data. Its direct-validation tools can encompass predictive, distributional, and substantive objects in our terminology, but these are not treated as three use-dependent dimensions. The two frameworks are therefore complementary and organized along different axes.
\\

\addlinespace

\citet{xie2026statistical}
&
Three cross-sectional and four longitudinal social-science datasets across multiple domains; synthetic data generated by 15 LLMs and evaluated through SSDataBench
&
Develops a systematic benchmark centered on \emph{population-level statistical realism}. Five types of statistical patterns are evaluated: univariate distributions, bivariate associations, multivariate outcome predictions, life-event sequence distributions, and associations between life-event sequences and covariates. Synthetic and real statistics are compared through repeated bootstrap-based significance tests summarized by pass rates. The framework explicitly argues that individual-level predictive performance should not be treated as the central validity criterion for population-oriented synthetic social-science data.
&
Provides the closest recent systematic treatment of population-level statistical realism. Univariate distributions correspond closely to our \emph{distributional validity}, while bivariate associations, multivariate relationships, and sequence--covariate associations overlap with objects that our framework places under \emph{substantive validity}. The organizing principle nevertheless differs: SSDataBench gives conceptual priority to population-level realism, whereas our framework retains paired respondent-level correspondence, marginal distributional fidelity, and preservation of downstream analysis-relevant relationships as distinct dimensions whose relative importance depends on the intended use of the synthetic survey sample.
\\

\addlinespace

\citet{park2026llms}
&
Korean Innovation Survey; GPT-4o-generated synthetic firm responses conditioned on observed firm characteristics
&
Develop a multi-dimensional validation framework combining qualitative consistency, quantitative replication of empirical patterns, the quality and contextual adequacy of generated explanations, and robustness to parameter variation. The quantitative exercises examine whether synthetic responses reproduce firm-size gradients and sectoral patterns associated with established innovation theory, while the qualitative exercises assess whether generated explanations are coherent with firm characteristics and theoretical expectations.
&
Overlaps most closely with \emph{substantive validity}, while comparisons of response levels and distributions also relate to \emph{distributional validity}. The framework is explicitly multidimensional, but its organizing dimensions combine quantitative pattern replication, qualitative explanation quality, and robustness. Our framework instead separates paired respondent-level correspondence, marginal distributional fidelity, and preservation of analysis-relevant relationships according to the intended empirical use of the synthetic survey sample.
\\

\addlinespace

\citet{cummins2026threat}
&
Silicon samples generated under alternative model, sampling, prompt, and respondent-information configurations; social-psychological scales and a reanalysis of an existing silicon-sample application
&
Examine the consequences of analytic flexibility in constructing silicon samples. In the first study, 252 alternative configurations are evaluated according to their ability to recover participant rankings, response distributions, and correlations between psychological scales. A second study varies the generation configuration in a published silicon-sample application and compares the resulting association structures. Performance varies substantially across generation choices and across evaluation criteria.
&
Spans objects related to \emph{predictive}, \emph{distributional}, and \emph{substantive validity} and provides particularly relevant evidence that good performance on one criterion need not imply good performance on another. Its central contribution concerns the sensitivity of conclusions to defensible generation choices, whereas our framework focuses on selecting distinct validation objects according to the intended empirical use of the synthetic sample.
\\

\end{longtable}

\end{landscape}

\section{Survey items, coding, and outcome construction}
\label{app:outcomes}

\setcounter{table}{0}
\renewcommand{\thetable}{B\arabic{table}}
\setcounter{figure}{0}
\renewcommand{\thefigure}{B\arabic{figure}}

This section documents the survey items, coding rules, sample construction, and empirical diagnostics used to construct the three outcome indices analysed in the paper. The harmonized item numbering Q01--Q18 follows the ordering used throughout the main paper and corresponds to the three outcomes retained from the original LLM generation procedure. The three outcomes are \emph{General} (Q01--Q05), \emph{Effects} (Q06--Q10), and \emph{Use} (Q11--Q18). The grouping follows the substantive organization of Special Eurobarometer 554 and is complemented by item-correlation and principal-component diagnostics.

\subsection*{B.1 Survey items}

Table~\ref{tab:appendix_outcome_items} reports the full English wording of the 18 survey items used to construct the three outcome indices. The wording follows the Eurobarometer questionnaire, while the Q01--Q18 identifiers are our harmonized labels rather than the original questionnaire variable names.

\begin{longtable}{p{1.15cm} p{13.0cm}}
\caption{Survey items used to construct the outcome indices}
\label{tab:appendix_outcome_items}\\
\toprule
\textbf{Code} & \textbf{Question} \\
\midrule
\endfirsthead

\multicolumn{2}{l}{\textit{Table \thetable\ continued}}\\
\toprule
\textbf{Code} & \textbf{Question} \\
\midrule
\endhead

\midrule
\multicolumn{2}{r}{\textit{Continued on next page}}\\
\endfoot

\bottomrule
\endlastfoot

\multicolumn{2}{l}{\textbf{General attitudes}}\\
\addlinespace
Q01 & In your view, what impact do the most recent digital technologies, including Artificial Intelligence, currently have on the economy? \\
Q02 & In your view, what impact do the most recent digital technologies, including Artificial Intelligence, currently have on society? \\
Q03 & In your view, what impact do the most recent digital technologies, including Artificial Intelligence, currently have on your quality of life? \\
Q04 & In your view, what impact do the most recent digital technologies, including Artificial Intelligence, currently have on social security benefits? \\
Q05 & To what extent do you agree or disagree with the following statement: Robots and Artificial Intelligence are a good thing for society, because they help people do their jobs or carry out daily tasks at home. \\

\addlinespace
\multicolumn{2}{l}{\textbf{Effects on work}}\\
\addlinespace
Q06 & How positively or negatively do you perceive the use of robots and Artificial Intelligence in the workplace? \\
Q07 & To what extent do you agree or disagree with the following statement: Artificial Intelligence is necessary as it can do jobs that are seen as boring or repetitive. \\
Q08 & To what extent do you agree or disagree with the following statement: Robots and Artificial Intelligence increase the pace at which workers complete tasks. \\
Q09 & To what extent do you agree or disagree with the following statement: Robots and Artificial Intelligence can be used to make accurate decisions in the workplace. \\
Q10 & To what extent do you agree or disagree with the following statement: Robots and Artificial Intelligence should be used more widely outside the workplace. \\

\addlinespace
\multicolumn{2}{l}{\textbf{Use at work}}\\
\addlinespace
Q11 & In general, how positively or negatively do you perceive the use of digital technologies, including Artificial Intelligence, in the workplace for gathering additional information on applicants for a job? \\
Q12 & In general, how positively or negatively do you perceive the use of digital technologies, including Artificial Intelligence, in the workplace for selecting applicants for a job? \\
Q13 & In general, how positively or negatively do you perceive the use of digital technologies, including Artificial Intelligence, in the workplace for allocating tasks to workers or managing their working schedules and shifts? \\
Q14 & In general, how positively or negatively do you perceive the use of digital technologies, including Artificial Intelligence, in the workplace for collecting, processing, and storing workers' personal data? \\
Q15 & In general, how positively or negatively do you perceive the use of digital technologies, including Artificial Intelligence, in the workplace for improving workers' safety and security? \\
Q16 & In general, how positively or negatively do you perceive the use of digital technologies, including Artificial Intelligence, in the workplace for monitoring workers? \\
Q17 & In general, how positively or negatively do you perceive the use of digital technologies, including Artificial Intelligence, in the workplace for assessing workers' performance? \\
Q18 & In general, how positively or negatively do you perceive the use of digital technologies, including Artificial Intelligence, in the workplace for automatically firing workers? \\

\end{longtable}

\noindent\textit{Notes:} The table reports the English wording of the survey items used in the analysis. The Q01--Q18 codes are harmonized identifiers used in this paper. The original questionnaire groups several items within common question batteries. All substantive response scales contain four ordered categories; non-substantive responses such as ``Don't know'' are treated as missing.

\subsection*{B.2 Coding, aggregation, and analytical samples}

The original response categories are harmonized to a common 1--4 analytical scale before index construction. For \emph{General}, \emph{Effects}, and \emph{Use}, higher values indicate more favorable evaluations of AI and digital technologies. Items whose original questionnaire coding runs from positive to negative are therefore reversed. Each composite outcome is then calculated as the unweighted arithmetic mean of its constituent items. The use of simple averages preserves the original 1--4 metric and avoids introducing outcome-specific weights into the comparison between observed and synthetic data.

Table~\ref{tab:appendix_index_construction} summarizes the construction of the three indices and distinguishes the pre-covariate samples from the final analytical samples used for the comparisons in the main paper.

\begin{table}[H]
\centering
\small
\caption{Outcome construction and analytical samples}
\label{tab:appendix_index_construction}
\renewcommand{\arraystretch}{1.15}
\setlength{\tabcolsep}{5pt}
\begin{tabular}{lcccc}
\toprule
\textbf{Index} & \textbf{Items} & \textbf{No. items} & \textbf{Pre-covariate N} & \textbf{Final analytical N} \\
\midrule
General & Q01--Q05 & 5 & 11,114 & 10,935 \\
Effects & Q06--Q10 & 5 & 12,843 & 12,583 \\
Use & Q11--Q18 & 8 & 12,843 & 12,583 \\
\bottomrule
\end{tabular}

\vspace{0.12cm}

\begin{minipage}{0.96\textwidth}
\footnotesize
\textit{Notes:} Pre-covariate N refers to the outcome-specific samples defined during data preparation after restricting the sample to respondents aged 18--65 and excluding retired respondents. \emph{Effects} and \emph{Use} use the common work-related complete-case sample, while \emph{General} is defined on the nested sample with complete Q01--Q05 responses. Final analytical N additionally requires complete information on the respondent characteristics used in the LLM profiles and supervised benchmarks.
\end{minipage}
\end{table}

The common work-related complete-case sample contains 12,843 respondents for \emph{Effects} and \emph{Use}, while the nested \emph{General} sample contains 11,114 respondents. Applying the respondent-information requirements used for the LLM and supervised-ML comparisons reduces the samples to 12,583 observations for \emph{Effects} and \emph{Use} and 10,935 for \emph{General}. Observed outcomes, synthetic outcomes, and supervised predictions are evaluated on these identical outcome-specific final samples.

\subsection*{B.3 Item-correlation structure}

\begin{figure}[H]
\centering
\includegraphics[width=\textwidth]{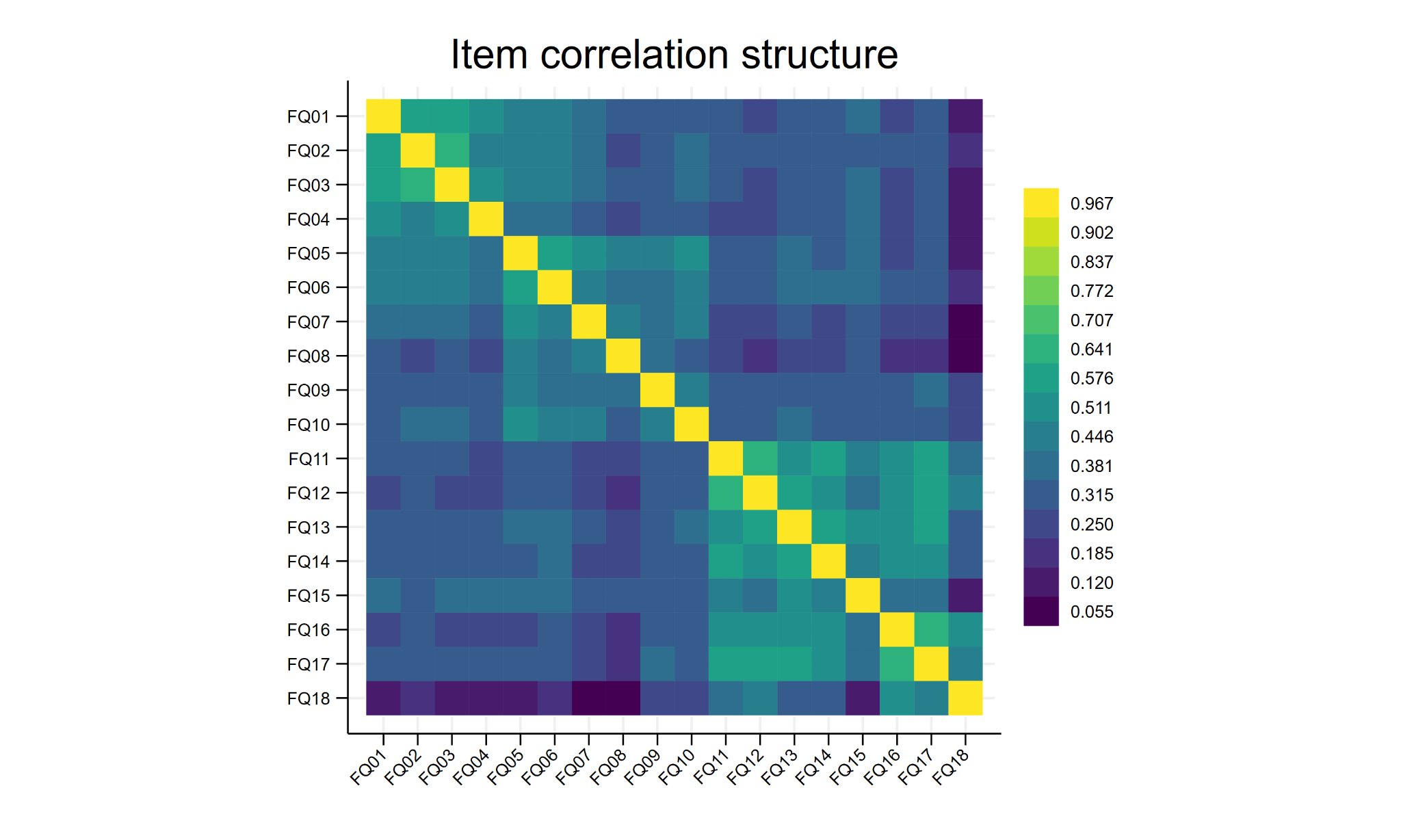}
\caption{Item-level correlation structure of the selected outcome items}
\label{fig:appendix_heatmap_final_pairwise}
\vspace{0.10cm}
\begin{minipage}{0.95\textwidth}
\footnotesize
\textit{Notes:} The figure reports pairwise Pearson correlations among the selected items after applying the missing-value rules used for index construction. Q01--Q05 form the \emph{General} block, Q06--Q10 the \emph{Effects} block, and Q11--Q18 the \emph{Use} block. Correlations are computed pairwise. All three outcome blocks are coded so that higher values indicate more favorable attitudes toward AI and digital technologies.
\end{minipage}
\end{figure}

The grouping of items is primarily substantive, reflecting the organization and content of the Eurobarometer questionnaire, but we also examine whether the selected items display the expected empirical structure. Figure~\ref{fig:appendix_heatmap_final_pairwise} reports pairwise Pearson correlations among Q01--Q18 after applying the item-level missing-value rules. The three outcome blocks remain empirically distinguishable and items within the same block generally display positive within-block correlations. Q05 is related to both broad attitudes and work-related evaluations, which is consistent with its wording combining a general evaluation of AI with references to work and everyday tasks; conceptually, we retain it in \emph{General}.

\subsection*{B.4 Principal-component diagnostics}

As an additional diagnostic, we conduct principal component analysis separately within each of the three item blocks. PCA is not used to define the final outcomes; the indices remain unweighted means. Its purpose is instead to assess whether the items assigned to each conceptual block load coherently on a common underlying component. Table~\ref{tab:appendix_pca_summary} shows that the first component has an eigenvalue greater than one for each block and explains between 51.85\% and 59.71\% of the within-block variance. All items have positive loadings on the first component in the coding direction used to construct the corresponding index.

\begin{table}[H]
\centering
\small
\caption{Principal-component diagnostics by outcome index}
\label{tab:appendix_pca_summary}
\renewcommand{\arraystretch}{1.15}
\setlength{\tabcolsep}{5pt}
\begin{tabular}{lcccc}
\toprule
\textbf{Index} & \textbf{Observations} & \textbf{Items} & \textbf{First eigenvalue} & \textbf{Variance explained by PC1} \\
\midrule
General & 11,114 & 5 & 2.985 & 59.71\% \\
Effects & 12,843 & 5 & 2.592 & 51.85\% \\
Use & 12,843 & 8 & 4.360 & 54.50\% \\
\bottomrule
\end{tabular}

\vspace{0.12cm}

\begin{minipage}{0.96\textwidth}
\footnotesize
\textit{Notes:} PCA is performed separately for each item block using the correlation matrix. Loadings on the first component are positive for all selected items: 0.390--0.476 for \emph{General}, 0.421--0.466 for \emph{Effects}, and 0.282--0.387 for \emph{Use}.
\end{minipage}
\end{table}

Taken together, the item-level correlations and PCA diagnostics support the three conceptual groupings used in the analysis. These diagnostics are not intended to establish that each block is unidimensional in a strict psychometric sense or to determine the index weights mechanically. Rather, they provide an empirical check that the questionnaire-based groupings used to construct the simple-average indices capture coherent response patterns while preserving a directly interpretable common 1--4 scale.

\section{Respondent profiles, prompting, and generation robustness}
\label{app:llm_generation}

\setcounter{table}{0}
\renewcommand{\thetable}{C\arabic{table}}

This section provides additional documentation on the respondent information supplied to the LLMs, the common prompting structure, the generation procedure, and the robustness of the synthetic outcomes to alternative temperature settings. The objective is to document the substantive information available to the models and the main generation choices without relying on provider-specific implementation details that are not relevant to the empirical comparison.

\subsection*{C.1 Respondent information}

Each synthetic respondent is conditioned on characteristics observed for the corresponding Eurobarometer respondent. The same substantive information base is subsequently used in tabular form by the supervised machine-learning benchmarks. No observed response entering the \emph{General}, \emph{Effects}, or \emph{Use} outcomes, and no index constructed from those responses, is included in the respondent profile.

Table C1 summarizes the respondent information represented in the textual profiles. It provides a conceptual description of the variables supplied to the models; the respondent-specific context template actually used in the generation procedure, including the textual definitions of categorical and constructed variables, is reproduced in Section~C.2.

\begin{longtable}{p{3.7cm} p{10.5cm}}
\caption{Summary of respondent information included in the LLM profiles}
\label{tab:profile_variables}\\
\toprule
\textbf{Profile component} & \textbf{Information supplied to the LLM} \\
\midrule
\endfirsthead

\multicolumn{2}{l}{\textit{Table \thetable\ continued}}\\
\toprule
\textbf{Profile component} & \textbf{Information supplied to the LLM} \\
\midrule
\endhead

\midrule
\multicolumn{2}{r}{\textit{Continued on next page}}\\
\endfoot

\bottomrule
\endlastfoot

Country & Country in which the respondent was interviewed. \\

Age & Respondent's age in years. \\

Gender & Respondent's reported gender category. \\

Community type & Type of community in which the respondent lives: rural area or village, small or medium-sized town, or large town. \\

Education & Four-category measure of educational attainment, represented through the corresponding substantive category. \\

Foreign background & Indicator of whether the respondent reports a nationality different from the country in which the interview was conducted. \\

Household size & Household size grouped into one, two, three, or four or more persons. \\

Children in household & Indicator of whether at least one child younger than 15 lives in the household. \\

Occupation & Broad occupational category, distinguishing self-employed workers, managers, other white-collar workers, manual workers, house persons, unemployed respondents, and students. \\

Employment status & Broad labour-market status distinguishing self-employed, employed, and not currently working respondents. \\

Sector of activity & Extended sector classification covering agriculture, manufacturing, logistics, services, public-sector activities, other sectors, and categories for respondents outside current employment. \\

Workplace size & Size of the respondent's workplace, ranging from one worker to more than 250 workers, with a separate category for respondents without a current workplace. \\

Exposure to AI at work & Measure of exposure to AI and digital technologies in workplace management activities, including hiring, task allocation, scheduling, monitoring, performance assessment, and safety-related applications. \\

Financial difficulties & Self-reported frequency of difficulties paying bills, represented as an ordered measure of financial conditions. \\

Life satisfaction & Respondent's reported level of overall life satisfaction. \\

Internet use & Three-category measure of the intensity of Internet use. \\

Political interest & Ordered measure of political engagement based on the reported frequency of political discussion. \\

View of the European Union & Ordered measure capturing the respondent's overall evaluation of the European Union. \\

Institutional trust & Measure capturing the respondent's assessment of the fairness and impartiality with which laws and rules are applied in the respondent's country. \\

\end{longtable}

The profile variables span demographic background, household composition, labour-market position, direct exposure to workplace technologies, economic conditions, general attitudes, and institutional context. They are intended to provide a relatively rich description of each respondent without incorporating information derived from the target AI-attitude questions themselves.

In the analyses, the measures described above as “View of the European Union” and “Institutional trust” are used as proxies for trust in EU institutions and trust in national institutions, respectively; these shorter labels are used in the machine-learning and predictor-ranking tables below.

\subsection*{C.2 Common prompting structure}

The substantive prompting strategy is held constant across the four model providers. Each generation consists of a system instruction, a respondent-specific textual profile, the complete set of survey items with their admissible response categories, and an instruction to return structured numerical answers. The system instruction used in the generation workflow is:

\begin{quote}
\small
\texttt{You simulate opinions of real individuals.}\\
\texttt{You always answer using a 1-4 scale.}\\
\texttt{Respond in clean JSON format with question IDs as keys and numeric responses as values.}\\
\texttt{Use only the provided individual attributes.}\\
\texttt{Keep responses minimal and structured.}
\end{quote}

The respondent-specific portion of the user prompt was generated from the common context template reproduced below. Square brackets indicate fields replaced by the corresponding observed respondent information. Importantly, the template supplied the models not only with the respondent-specific category but also, where relevant, with textual definitions of the categories and constructed variables.

\begin{quote}
\small
You have been interviewed from \textbf{[country]} and have reported the following information about yourself in that interview:

You are a \textbf{[age]} years old \textbf{[gender]} living in a \textbf{[community type]} (which you chose among these alternatives: Rural area or village, Small or middle sized town, Large town).

Citizenship status: \textbf{[foreign background]}

Your educational background: \textbf{[education]}

Household composition: \textbf{[household size]}, with \textbf{[children status]}

Your work and occupation:

Occupation: \textbf{[occupation]} (which you chose among these alternatives: Self-employed [self-employed farmer, self-employed fisherman, self-employed professional, owner of a shop or crafts business, business proprietor or company owner]; Managers [employed professionals, general or top management, middle management, department heads, teachers and technicians]; Other white collars [desk-based employees and travelling employees]; Manual workers [service workers, supervisors, skilled manual workers and other manual workers]; House persons; Unemployed; Students).

Employment status: \textbf{[employment status]}

Sector of activity: \textbf{[sector]} (the statement ``Other'' excludes Agriculture, Manufacturing, Logistics, Services, Public sector, House persons, Unemployed, and Students).

Workplace size by number of persons: \textbf{[workplace size]}

Exposure to AI at work: \textbf{[AI exposure]} (High means being exposed on average in the workplace to AI through hiring, task allocation, schedule management, activity monitoring, performance assessment including rewards or sanctions, and safety enforcement; Low means having low average exposure in the workplace to AI through these practices; Never worked implies no prior exposure to AI at work).

Your economic situation and well-being:

Financial difficulties: \textbf{[financial difficulties]} (which you chose among the following alternatives: Most of the time, From time to time, Almost never/never).

Life satisfaction: \textbf{[life satisfaction]} (which you chose among these alternatives: Very satisfied, Fairly satisfied, Not very satisfied, Not at all satisfied).

Internet use: \textbf{[Internet use]} (which you chose among these alternatives: Daily users, Non-daily users, Never/no access).

Your political engagement:

Political interest: \textbf{[political interest]} (which you chose among these alternatives: Strong, Medium, Low, Not at all).

View of the European Union: \textbf{[view of the European Union]} (which you chose among these alternatives: Very positive, Fairly positive, Neutral, Fairly negative, Very negative).

Trust in how law is applied in your country: \textbf{[institutional trust]} (which you chose among these alternatives: Totally agree, Tend to agree, Tend to disagree, Totally disagree).

I am going to ask you the following questions and you answer them for me:
\end{quote}

In the original generation workflow, the respondent profile was followed by the complete set of questionnaire items used at that stage, identified as Q01--Q24. The final analysis in this paper retains Q01--Q18, corresponding to the \emph{General}, \emph{Effects}, and \emph{Use} outcomes reported in Table~B1; the additional items Q19--Q24 were part of the same generation request but are not analysed in the final paper. The generation request concludes with the instruction:

\begin{quote}
\small
\texttt{Please respond in JSON format with question IDs as keys.}
\end{quote}

No examples of desirable responses, observed population frequencies, group means, observed outcome values, or other target-specific calibration information are provided. All questions are submitted together for each respondent, so the model produces a single item-level response vector under a common respondent context.

\subsection*{C.3 Models, generation settings, and output processing}

Table~\ref{tab:llm_settings} summarizes the models and generation settings used in the main analysis. All four models are used off the shelf, without task-specific fine-tuning or supervised calibration on the Eurobarometer outcomes.

\begin{table}[H]
\centering
\small
\caption{LLMs and main generation settings}
\label{tab:llm_settings}
\renewcommand{\arraystretch}{1.12}
\setlength{\tabcolsep}{6pt}
\begin{tabular}{lll}
\toprule
\textbf{Model label} & \textbf{Model} & \textbf{Main temperature} \\
\midrule
Claude & Claude Sonnet 4.6 & 0.5 \\
GPT & GPT-4.1 mini & 0.5 \\
Gemini & Gemini 3 Flash & 0.5 \\
Grok & Grok 4.1 Fast Non-Reasoning & 0.5 \\
\bottomrule
\end{tabular}

\vspace{0.12cm}

\begin{minipage}{0.94\textwidth}
\footnotesize
\textit{Notes:} The same substantive prompt structure and respondent-information base are used across providers. The nominal temperature setting \(T=0.5\) is used for the main synthetic samples; because temperature is implemented on provider-specific scales, the same nominal value should not be interpreted as implying identical effective randomness across models. GPT, Gemini, and Grok are additionally evaluated at \(T=0.2\) and \(T=1.0\) in the robustness exercise below. Claude was not regenerated at the alternative temperatures because repeating the complete extraction involved substantially higher computational cost.
\end{minipage}
\end{table}

Generated outputs are parsed as question--response pairs. A response is retained only when it can be interpreted as a numerical value within the admissible 1--4 range; invalid or unparseable responses are treated as missing. In the retained synthetic datasets, all item-level outputs were successfully parsed and fell within the admissible 1--4 range, so this rule did not generate any missing synthetic responses. Valid generated responses are not otherwise adjusted, rescaled, or calibrated. The original generation workflow produced responses for Q01--Q24. The final analysis reported in this paper retains Q01--Q18, which are processed using the same coding direction and simple-average index construction applied to the observed survey data.

\subsection*{C.4 Robustness to temperature}

To assess whether the main findings depend materially on the common \(T=0.5\) specification, complete synthetic samples were regenerated for GPT, Gemini, and Grok at \(T=0.2\) and \(T=1.0\). We focus on the three core outcomes used in the main analysis. Table~\ref{tab:temperature_descriptives} reports the mean and standard deviation of each synthetic index across temperature settings.

\begin{table}[H]
\centering
\scriptsize
\caption{Synthetic index distributions across temperature settings}
\label{tab:temperature_descriptives}
\renewcommand{\arraystretch}{1.08}
\setlength{\tabcolsep}{4.2pt}
\begin{tabular}{llcccccc}
\toprule
& & \multicolumn{2}{c}{\textbf{General}} & \multicolumn{2}{c}{\textbf{Effects}} & \multicolumn{2}{c}{\textbf{Use}} \\
\cmidrule(lr){3-4}\cmidrule(lr){5-6}\cmidrule(lr){7-8}
\textbf{Model} & \textbf{Temperature} & \textbf{Mean} & \textbf{SD} & \textbf{Mean} & \textbf{SD} & \textbf{Mean} & \textbf{SD} \\
\midrule
GPT & 0.2 & 2.870 & 0.491 & 2.883 & 0.466 & 2.109 & 0.531 \\
GPT & 0.5 & 2.868 & 0.496 & 2.884 & 0.475 & 2.105 & 0.530 \\
GPT & 1.0 & 2.865 & 0.514 & 2.890 & 0.492 & 2.096 & 0.535 \\
\addlinespace
Gemini & 0.2 & 2.880 & 0.450 & 2.957 & 0.423 & 2.048 & 0.360 \\
Gemini & 0.5 & 2.883 & 0.453 & 2.961 & 0.426 & 2.047 & 0.360 \\
Gemini & 1.0 & 2.899 & 0.450 & 2.982 & 0.416 & 2.045 & 0.366 \\
\addlinespace
Grok & 0.2 & 2.703 & 0.767 & 2.397 & 0.580 & 1.693 & 0.368 \\
Grok & 0.5 & 2.703 & 0.773 & 2.391 & 0.582 & 1.691 & 0.367 \\
Grok & 1.0 & 2.703 & 0.771 & 2.397 & 0.586 & 1.695 & 0.369 \\
\bottomrule
\end{tabular}

\vspace{0.12cm}

\begin{minipage}{0.97\textwidth}
\footnotesize
\textit{Notes:} Entries report the mean and standard deviation of the complete synthetic indices generated at each temperature. The \(T=0.5\) samples correspond to the main analysis.
\end{minipage}
\end{table}

The marginal distributions are highly stable across the three temperature settings. The largest absolute difference in an index mean relative to \(T=0.5\) is approximately 0.02 points on the 1--4 outcome scale, while synthetic standard deviations also change only modestly. Temperature therefore does not materially alter the level or dispersion patterns that characterize the main synthetic samples.

Table~\ref{tab:temperature_correlations} complements this comparison by matching respondents across temperature runs and reporting correlations between the alternative-temperature synthetic indices and their \(T=0.5\) counterparts.

\begin{table}[H]
\centering
\small
\caption{Respondent-level stability across temperature settings}
\label{tab:temperature_correlations}
\renewcommand{\arraystretch}{1.10}
\setlength{\tabcolsep}{7pt}
\begin{tabular}{llccc}
\toprule
\textbf{Model} & \textbf{Comparison with \(T=0.5\)} & \textbf{General} & \textbf{Effects} & \textbf{Use} \\
\midrule
GPT & \(T=0.2\) & 0.943 & 0.919 & 0.909 \\
GPT & \(T=1.0\) & 0.899 & 0.866 & 0.872 \\
\addlinespace
Gemini & \(T=0.2\) & 0.914 & 0.854 & 0.869 \\
Gemini & \(T=1.0\) & 0.902 & 0.838 & 0.851 \\
\addlinespace
Grok & \(T=0.2\) & 0.970 & 0.948 & 0.922 \\
Grok & \(T=1.0\) & 0.965 & 0.936 & 0.904 \\
\bottomrule
\end{tabular}

\vspace{0.12cm}

\begin{minipage}{0.94\textwidth}
\footnotesize
\textit{Notes:} Entries are Pearson correlations between synthetic respondent-level indices generated at the alternative temperature and the corresponding \(T=0.5\) index for the same respondents. Respondents are matched across runs using their survey identifier.
\end{minipage}
\end{table}

Respondent-level synthetic outcomes also remain strongly correlated across temperature specifications. Correlations range from 0.838 to 0.970, with the lower values generally occurring at \(T=1.0\), as expected from the greater stochasticity of generation. These changes are considerably larger at the individual-response level than the changes in aggregate means and standard deviations, but they do not alter the qualitative validation patterns reported in the main paper. In particular, the relative difficulties of the \emph{Use} outcome and the model-specific differences in calibration and dispersion remain visible across settings.

The robustness exercise therefore indicates that the principal findings are not an artifact of choosing \(T=0.5\). Alternative temperatures change particular respondent-level generations, but the aggregate properties relevant to the validation framework and the qualitative conclusions of the application remain stable.

\section{Supervised machine-learning implementation and reproducibility}
\label{app:ml_implementation}

\setcounter{table}{0}
\renewcommand{\thetable}{D\arabic{table}}

This section documents the predictor representation, preprocessing, country fixed-effect treatment, resampling design, and tuning procedures used to construct the supervised machine-learning benchmarks reported in the main paper. The objective of these models is to provide an out-of-sample supervised reference based on the same substantive respondent-information base available to the LLMs, rather than to construct a technically equivalent competitor.

\subsection*{D.1 Predictor representation and analytical samples}

The supervised models use the respondent characteristics represented in the LLM profiles, encoded in a tabular form suitable for statistical learning. Table~\ref{tab:ml_predictors} reports the variables entering the prediction models and their treatment in the preprocessing pipeline.

\begin{longtable}{p{4.0cm} p{4.0cm} p{6.2cm}}
\caption{Predictors used in the supervised machine-learning benchmarks}
\label{tab:ml_predictors}\\
\toprule
\textbf{Variable} & \textbf{Representation} & \textbf{Preprocessing} \\
\midrule
\endfirsthead

\multicolumn{3}{l}{\textit{Table \thetable\ continued}}\\
\toprule
\textbf{Variable} & \textbf{Representation} & \textbf{Preprocessing} \\
\midrule
\endhead

\midrule
\multicolumn{3}{r}{\textit{Continued on next page}}\\
\endfoot

\bottomrule
\endlastfoot

Age & Continuous & Numerical; standardized within the training data. \\

Community type & Ordered categorical & Entered numerically and standardized. \\

Education & Ordered categorical & Four-category education measure entered numerically and standardized. \\

Household size & Ordered categorical & Four-category household-size measure entered numerically and standardized. \\

Financial condition & Ordered categorical & Entered numerically and standardized. \\

Life satisfaction & Ordered categorical & Entered numerically and standardized. \\

Internet use & Ordered categorical & Three-category measure entered numerically and standardized. \\

Political interest & Ordered categorical & Entered numerically and standardized. \\

Trust in the EU & Ordered categorical & Entered numerically and standardized. \\

Institutional trust & Ordered categorical & Entered numerically and standardized. \\

Gender & Categorical & One-hot encoded. \\

Foreign background & Binary categorical & One-hot encoded. \\

Children in household & Binary categorical & One-hot encoded. \\

Exposure to AI at work & Categorical block & One-hot encoded. \\

Occupation & Categorical block & One-hot encoded. \\

Sector of activity & Categorical block & One-hot encoded. \\

Workplace size & Categorical block & One-hot encoded. \\

Country & Fixed effect & Represented by a complete set of country indicators and partialled out before model fitting. \\

\end{longtable}

No imputation of predictor values is performed. The common predictor sample is first restricted to respondents with complete information on the variables entering the prediction models and the country identifier. Outcome-specific samples are then obtained by additionally requiring a valid value of the corresponding observed index. The resulting samples contain \(N=10{,}935\) respondents for \emph{General} and \(N=12{,}583\) for \emph{Effects} and \emph{Use}.

\subsection*{D.2 Fold-specific preprocessing and country fixed effects}

For every final out-of-fold prediction, transformations needed to construct the predictor matrix are estimated using training observations only and then applied to the corresponding held-out observations. Categorical predictors are one-hot encoded, with the first category omitted, and previously unseen categories in a held-out fold are handled without modifying the training representation. Numerical and encoded predictors are subsequently standardized using means and standard deviations estimated on the training observations.

Country differences are handled through fixed-effect residualization. Let \(Z\) denote the complete set of country indicators. Within each training sample, both the standardized predictor matrix \(X\) and the target outcome \(Y\) are projected on \(Z\). The prediction models are then fitted to the residualized quantities. The country component of the target estimated from the training data is subsequently added back to the within-country prediction for each held-out respondent, yielding final predictions on the original 1--4 outcome scale.

This procedure serves two purposes. First, the learners focus on within-country relationships between respondent characteristics and outcomes rather than using country differences as an unrestricted source of predictive performance. Second, adding the estimated country component back after prediction allows the final supervised predictions to be compared directly with the observed and LLM-generated outcomes in levels.

\subsection*{D.3 Out-of-fold prediction and hyperparameter tuning}

All supervised results reported in the main paper use 10-fold out-of-fold predictions. The outcome-specific sample is divided into ten shuffled folds. Each respondent is predicted by a model estimated without that respondent, and the predictions from the ten held-out folds are subsequently recombined to form a full out-of-fold prediction vector. A common random seed of 12,345 is used throughout the resampling and model-fitting procedures.

The tuning strategy differs slightly across algorithms because of their computational requirements. Elastic Net uses fully fold-specific tuning: within each outer training fold, a second 10-fold cross-validation selects the regularization parameters before the fitted model predicts the held-out outer fold. Random Forest and Histogram Gradient Boosting are substantially more computationally intensive. Their hyperparameters are therefore selected once for each outcome through 10-fold cross-validation on the predefined original training sample and are then held fixed during the final 10-fold out-of-fold prediction procedure. Importantly, the final prediction for every respondent remains out of sample even when the tuning parameters are selected once per outcome.

Table~\ref{tab:ml_tuning} reports the parameter grids used for the three base learners.

\begin{table}[H]
\centering
\scriptsize
\caption{Hyperparameter tuning grids for the supervised benchmarks}
\label{tab:ml_tuning}
\renewcommand{\arraystretch}{1.16}
\setlength{\tabcolsep}{4.5pt}
\begin{tabular}{p{3.0cm} p{4.0cm} p{7.0cm}}
\toprule
\textbf{Model} & \textbf{Hyperparameter} & \textbf{Candidate values} \\
\midrule

Elastic Net & \(L_1\) ratio & \(0.5,\ 1.0\) \\

& Penalty parameter \(\alpha\) & 40 logarithmically spaced values between \(10^{-4}\) and \(1\) \\

\addlinespace

Random Forest & Number of trees & \(500,\ 1000\) \\

& Maximum tree depth & unrestricted, \(10,\ 20\) \\

& Minimum observations per leaf & \(1,\ 5,\ 10\) \\

& Features considered per split & square root of the number of features, \(0.5\) of features \\

\addlinespace

Histogram Gradient Boosting & Maximum tree depth & \(3,\ 5,\) unrestricted \\

& Learning rate & \(0.03,\ 0.06,\ 0.10\) \\

& Boosting iterations & \(300,\ 600\) \\

& Minimum observations per leaf & \(10,\ 30,\ 50\) \\

& \(L_2\) regularization & \(0,\ 0.1\) \\

\bottomrule
\end{tabular}

\vspace{0.12cm}

\begin{minipage}{0.97\textwidth}
\footnotesize
\textit{Notes:} Hyperparameters are selected by minimizing cross-validated mean squared error. Elastic Net tuning is performed separately inside each outer training fold. Random Forest and Histogram Gradient Boosting are tuned once per outcome on the predefined training sample and the selected parameters are subsequently held fixed during 10-fold out-of-fold prediction.
\end{minipage}
\end{table}

Elastic Net is estimated with a maximum of 100,000 optimization iterations and a convergence tolerance of \(10^{-6}\). Histogram Gradient Boosting uses squared-error loss. Random Forest uses all available processor cores during fitting. These implementation choices are held constant across outcomes.

\subsection*{D.4 Super Learner and prediction outputs}

The Super Learner combines Elastic Net, Random Forest, and Histogram Gradient Boosting through stacked regression. Outcome-specific base-learner parameters are first selected using the procedures described above. Within each outer training fold, the three base learners then generate the predictions used by a five-fold internal stacking procedure. The meta-learner is an Elastic Net with \(\alpha=0.001\) and \(L_1\) ratio \(0.1\). The final stacked prediction for each respondent is produced from an outer-fold model that does not include that respondent in its training sample.

For computational feasibility, the number of Random Forest trees used inside the Super Learner is capped at 500 even when the independently tuned Random Forest specification selects 1,000 trees. The remaining selected Random Forest and Histogram Gradient Boosting parameters are retained. This restriction affects only the Random Forest component inside the stacked model and not the standalone Random Forest benchmark.

The main validation analyses use the raw out-of-fold predictions reconstructed on the original outcome scale. Because these regression predictions are not mechanically constrained to the observed 1--4 support, the implementation also produces versions clipped to the interval \([1,4]\). These bounded predictions are retained only as a descriptive robustness check and are not used for the main results reported in the paper.

All preprocessing, resampling, fixed-effect residualization, tuning, and prediction procedures are implemented in Python using standard numerical and machine-learning libraries. The fixed seed of 12,345, the explicit parameter grids above, and the fold-specific fitting of preprocessing transformations make the prediction workflow reproducible conditional on the analytical data and software environment.

\section{Additional distributional evidence}
\label{app:distribution_plots}

\setcounter{figure}{0}
\renewcommand{\thefigure}{E\arabic{figure}}

The distributional-validity diagnostics reported in the main paper summarize differences in location, dispersion, and shape numerically. This section complements those measures with kernel-density plots of the observed outcomes, the four LLM-generated synthetic samples at \(T=0.5\), and the four supervised out-of-fold prediction distributions. The figures are descriptive rather than additional validation tests. In particular, the vertical density scale may differ between the LLM and ML panels, so peak heights should be interpreted within rather than across panels.

\begin{figure}[H]
\centering
\begin{subfigure}[t]{0.48\textwidth}
\centering
\includegraphics[width=\linewidth]{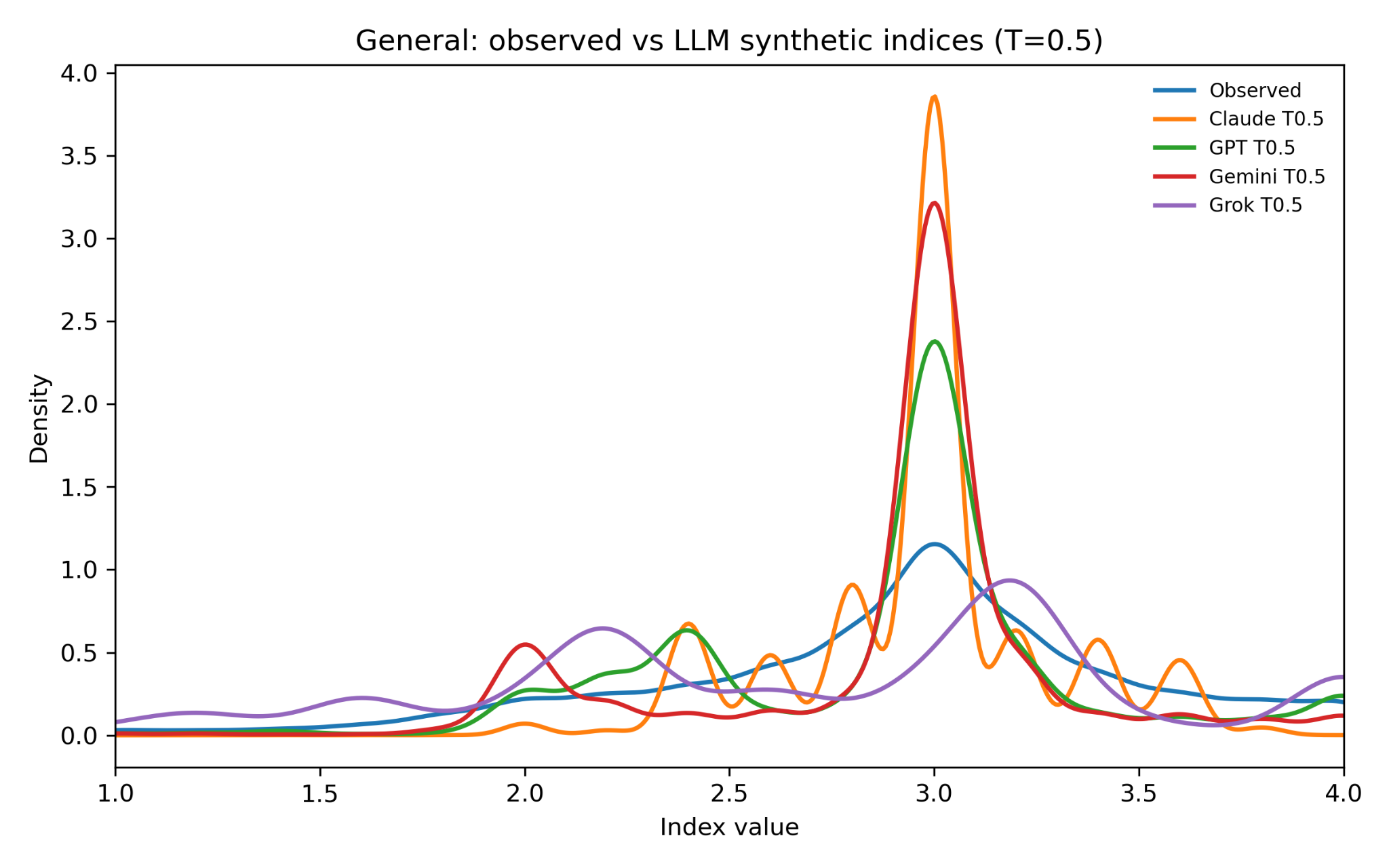}
\caption{Observed vs LLM-generated outcomes}
\end{subfigure}
\hfill
\begin{subfigure}[t]{0.48\textwidth}
\centering
\includegraphics[width=\linewidth]{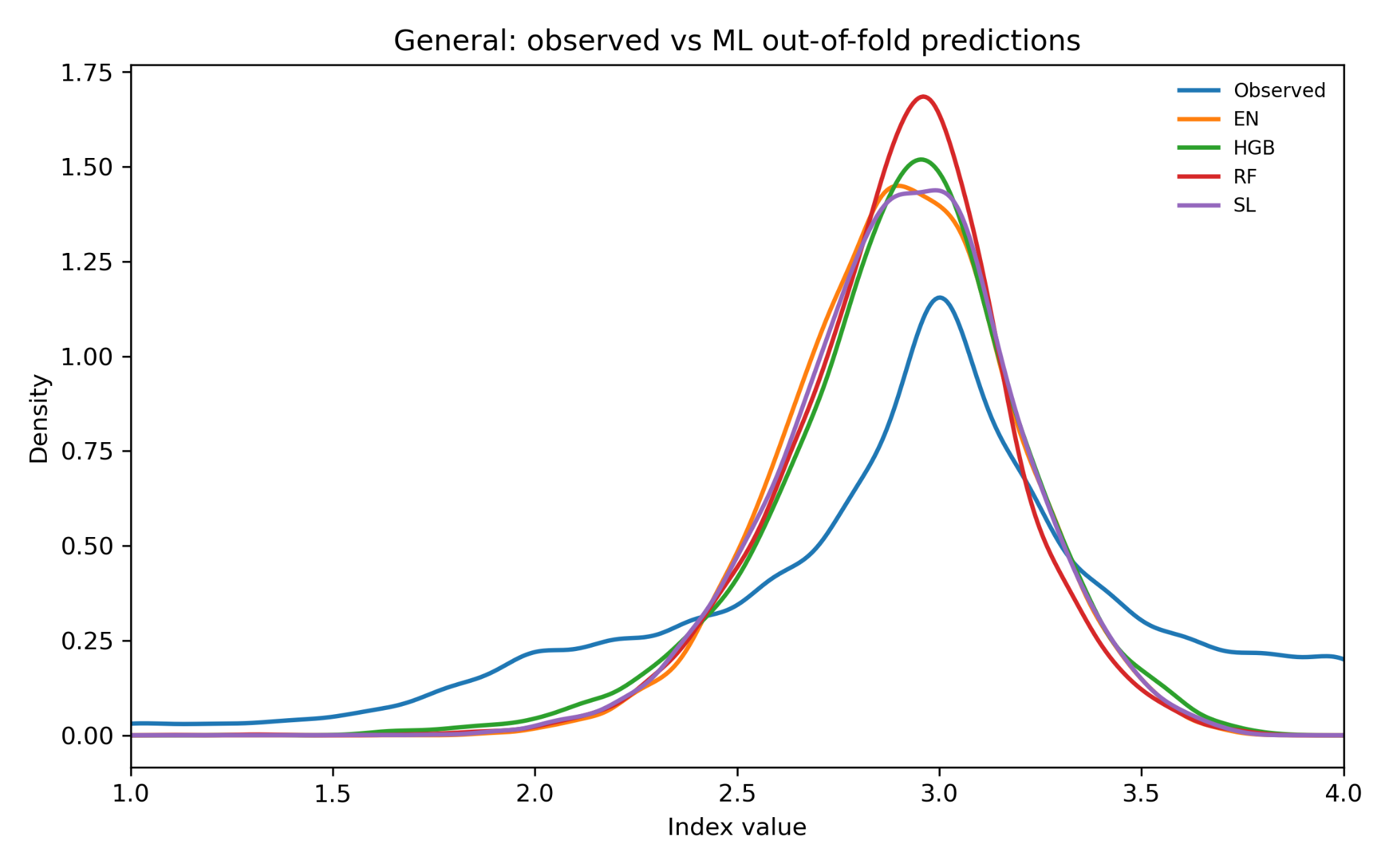}
\caption{Observed vs ML out-of-fold predictions}
\end{subfigure}
\caption{Distributional comparison for the \emph{General} index}
\label{fig:appendix_density_general}
\vspace{0.10cm}
\begin{minipage}{0.95\textwidth}
\footnotesize
\textit{Notes:} The left panel compares the observed \emph{General} distribution with the four LLM-generated synthetic distributions at \(T=0.5\). The right panel compares the observed distribution with the out-of-fold predictions from Elastic Net, Random Forest, Histogram Gradient Boosting, and Super Learner. All distributions refer to the common \emph{General} analytical sample.
\end{minipage}
\end{figure}

The figure illustrates the distinction emphasized by the formal distributional diagnostics. The supervised predictions are concentrated more strongly around the center of the observed distribution, while the LLM-generated samples display larger differences in dispersion and shape across models. GPT and Gemini preserve substantially more marginal variation than the supervised predictions, whereas Grok displays excess dispersion together with the mean miscalibration documented in the main paper.

\begin{figure}[H]
\centering
\begin{subfigure}[t]{0.48\textwidth}
\centering
\includegraphics[width=\linewidth]{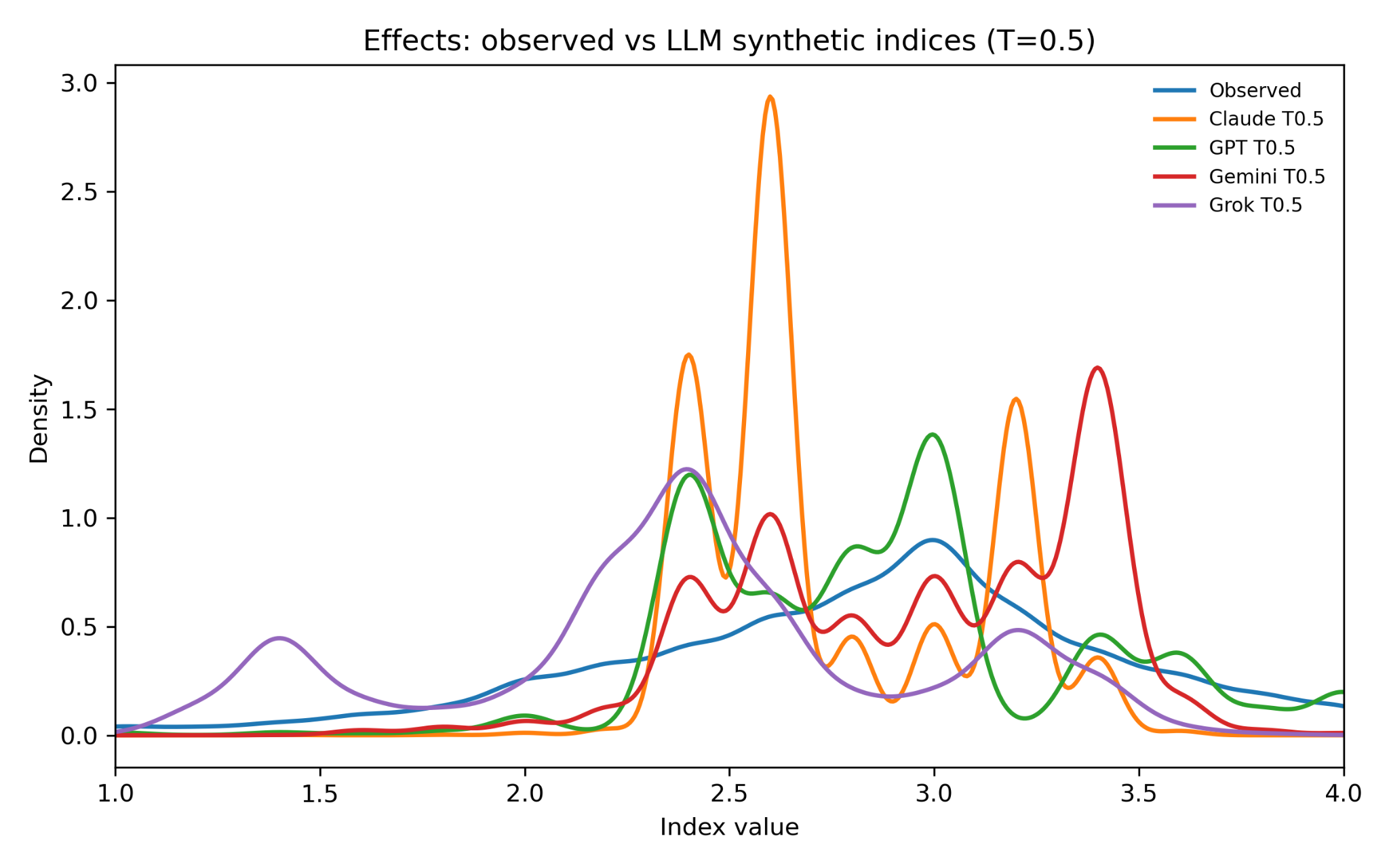}
\caption{Observed vs LLM-generated outcomes}
\end{subfigure}
\hfill
\begin{subfigure}[t]{0.48\textwidth}
\centering
\includegraphics[width=\linewidth]{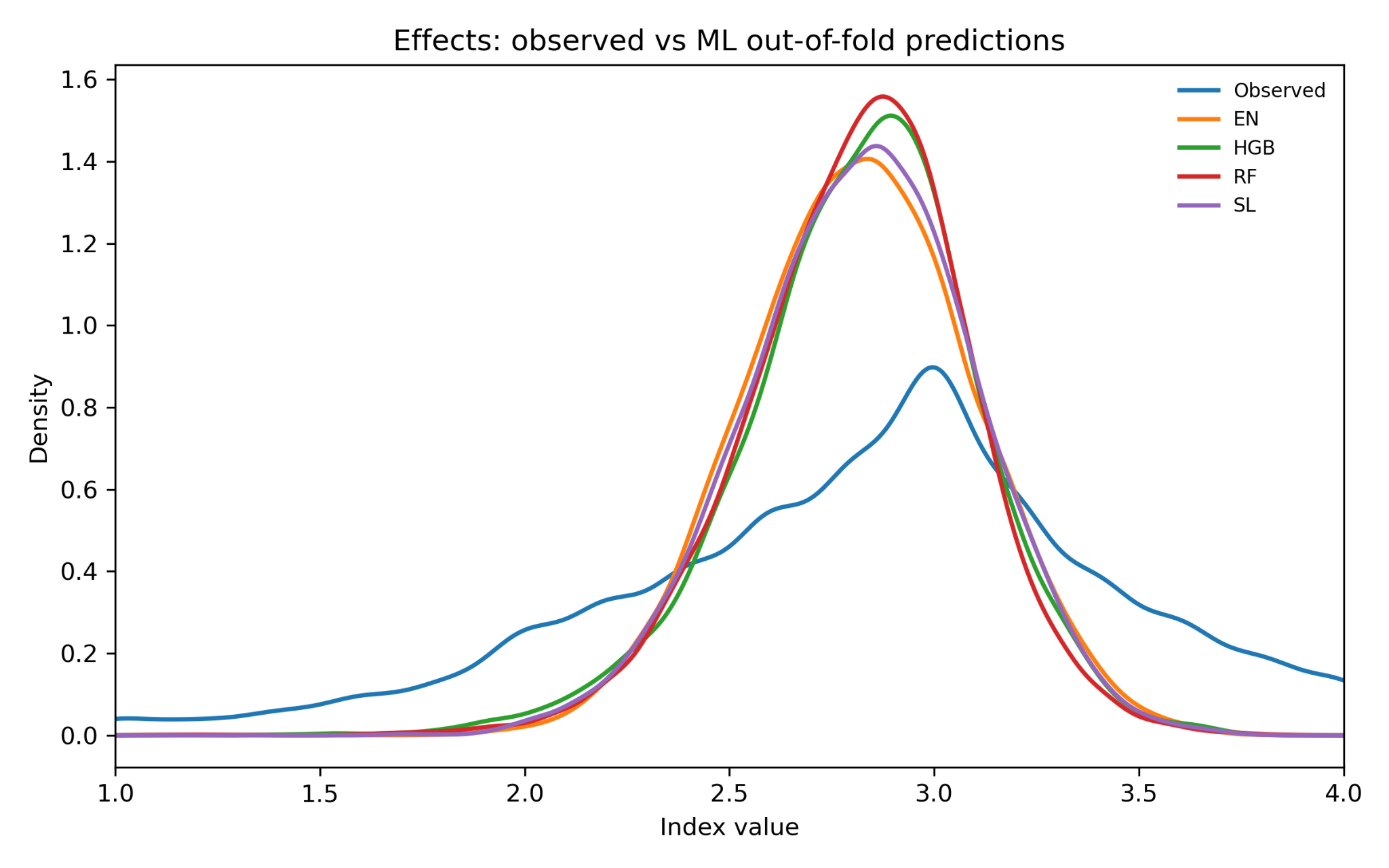}
\caption{Observed vs ML out-of-fold predictions}
\end{subfigure}
\caption{Distributional comparison for the \emph{Effects} index}
\label{fig:appendix_density_effects}
\vspace{0.10cm}
\begin{minipage}{0.95\textwidth}
\footnotesize
\textit{Notes:} The left panel compares the observed \emph{Effects} distribution with the four LLM-generated synthetic distributions at \(T=0.5\). The right panel compares the observed distribution with the four supervised out-of-fold prediction distributions. All distributions refer to the common \emph{Effects} analytical sample.
\end{minipage}
\end{figure}

For \emph{Effects}, the supervised predictions again show substantial compression relative to the observed outcome. Among the LLM-generated samples, GPT reproduces several aspects of the observed marginal shape relatively closely despite its weaker respondent-level predictive performance, whereas other models display larger shifts in location or concentration. The visual evidence is therefore consistent with the quantitative finding that predictive and distributional validity need not rank the same synthetic samples in the same way.

\begin{figure}[H]
\centering
\begin{subfigure}[t]{0.48\textwidth}
\centering
\includegraphics[width=\linewidth]{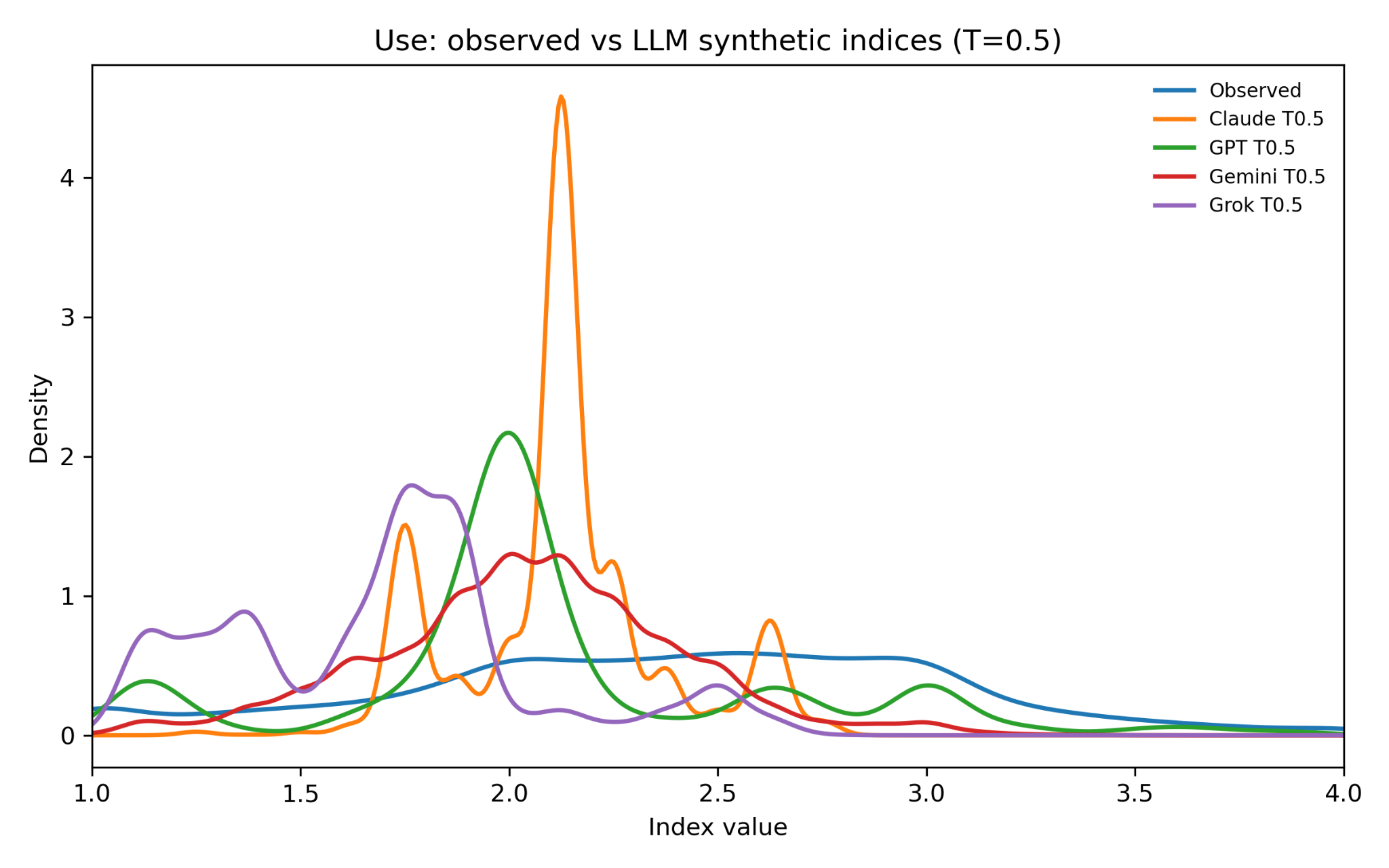}
\caption{Observed vs LLM-generated outcomes}
\end{subfigure}
\hfill
\begin{subfigure}[t]{0.48\textwidth}
\centering
\includegraphics[width=\linewidth]{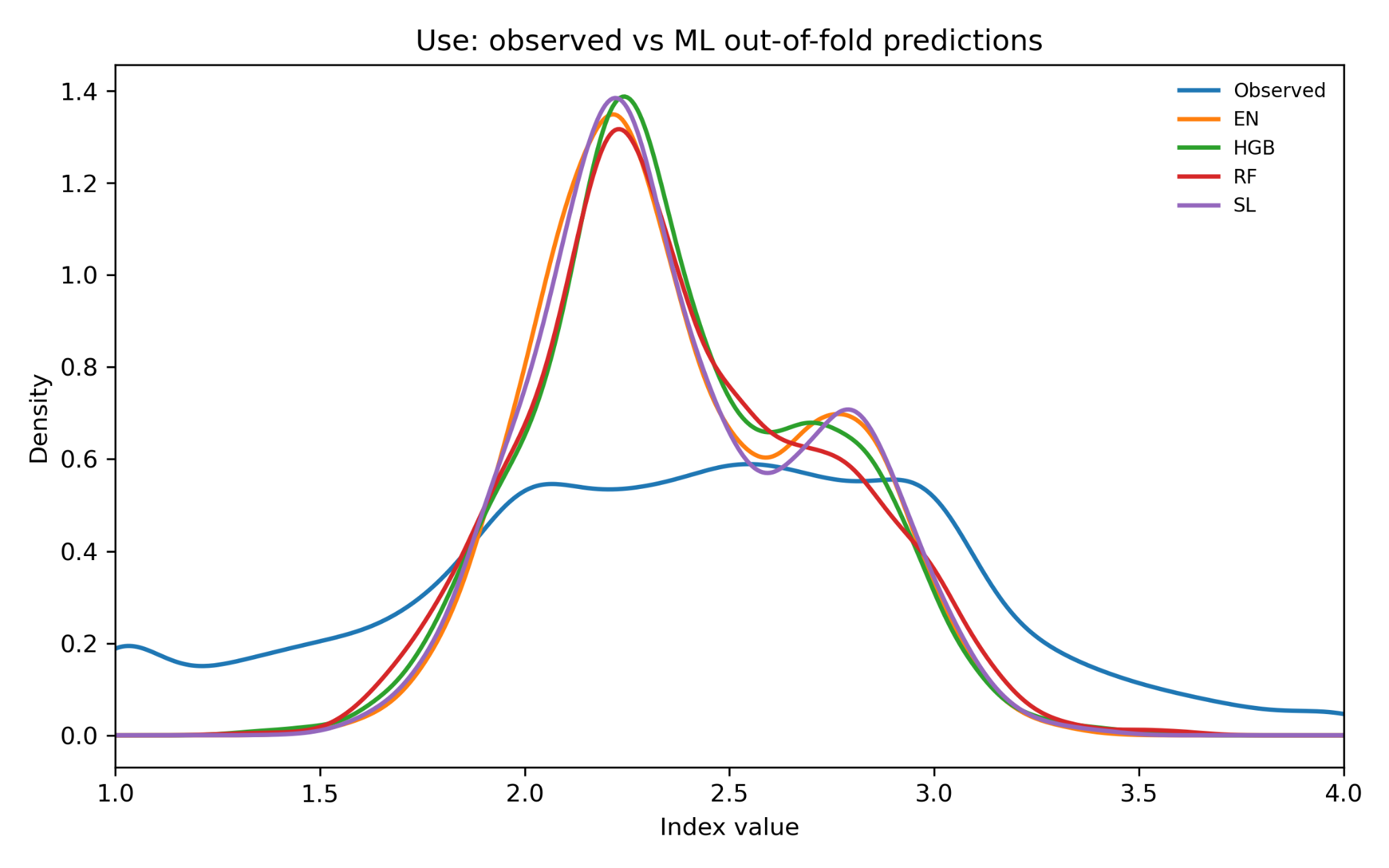}
\caption{Observed vs ML out-of-fold predictions}
\end{subfigure}
\caption{Distributional comparison for the \emph{Use} index}
\label{fig:appendix_density_use}
\vspace{0.10cm}
\begin{minipage}{0.95\textwidth}
\footnotesize
\textit{Notes:} The left panel compares the observed \emph{Use} distribution with the four LLM-generated synthetic distributions at \(T=0.5\). The right panel compares the observed distribution with the four supervised out-of-fold prediction distributions. All distributions refer to the common \emph{Use} analytical sample.
\end{minipage}
\end{figure}

The \emph{Use} comparison is less favorable for the LLM-generated samples. Several synthetic distributions are shifted toward lower values and fail to reproduce important features of the observed distribution. Although GPT retains more marginal dispersion than the other LLMs, this additional variation does not remove the negative mean bias and distributional discrepancies reported in the main analysis. The supervised predictions remain substantially compressed but are more closely calibrated in location.

Taken together, Figures~\ref{fig:appendix_density_general}--\ref{fig:appendix_density_use} visually reinforce the central distributional result of the paper. Supervised prediction can achieve close mean calibration while reproducing only part of the observed marginal heterogeneity, whereas LLM-generated samples may retain greater variation but also introduce model-specific distortions in location and shape. These plots therefore complement, rather than replace, the numerical distributional-validity diagnostics reported in the main paper.

\section{Complete predictor-structure rankings}
\label{app:predictor_rankings}

\setcounter{table}{0}
\renewcommand{\thetable}{F\arabic{table}}

This section reports the complete block-level predictor rankings underlying the substantive-validity comparison summarized in the main paper. The exercise is conducted ex post on the observed outcome and on the synthetic outcome vectors already generated by each LLM. No additional LLM calls are made and no respondent profiles are modified during the ablation analysis. Instead, Elastic Net is used as a common diagnostic model applied separately to the observed outcome and to each fixed synthetic outcome. For each source and outcome, the full model is first estimated using 10-fold out-of-fold prediction. One predictor block is then removed at a time and the model is re-estimated. A predictor block corresponds to one substantive predictor; for categorical predictors represented by multiple indicators, all indicators corresponding to that predictor are removed jointly. Block importance is defined as the increase in cross-validated mean squared error, \(\Delta\mathrm{MSE}=\mathrm{MSE}_{-G}-\mathrm{MSE}_{\mathrm{full}}\). Larger values therefore indicate that removing the block produces a larger deterioration in predictive performance.

The Elastic Net used for this diagnostic exercise is tuned within each outer training fold using five-fold internal cross-validation. The candidate \(L_1\) ratios are \(0.1\), \(0.5\), \(0.9\), and \(1.0\), and the penalty parameter is selected from 60 logarithmically spaced values between \(10^{-4}\) and \(10^{1}\). For each dropped-block model, the tuning parameters selected for the corresponding full-model outer fold are retained. The same covariates, analytical sample, predictor blocks, and ablation procedure are used for the observed and synthetic outcomes. Country fixed effects and predictor preprocessing follow the same fold-safe residualization logic described in Section~D. The resulting rankings are predictive rather than causal and measure whether the observed and synthetic outcomes imply similar empirical importance structures across respondent characteristics.

\subsection*{F.1 General}

Table~\ref{tab:ranking_general} reports the complete rankings for the \emph{General} outcome. Trust in EU institutions is ranked first for the observed outcome and for all four LLM-generated outcomes. The principal discrepancy among the highest-ranked observed predictors concerns AI and technology exposure at work, which ranks second in the observed data but substantially lower in the synthetic outcomes, particularly for Claude and Grok.

\begin{table}[H]
\centering
\scriptsize
\caption{Complete predictor rankings for the \emph{General} outcome}
\label{tab:ranking_general}
\renewcommand{\arraystretch}{1.08}
\setlength{\tabcolsep}{5pt}
\begin{tabular}{lccccc}
\toprule
\textbf{Predictor block} & \textbf{Observed} & \textbf{Claude} & \textbf{GPT} & \textbf{Gemini} & \textbf{Grok} \\
\midrule
Trust in EU institutions & 1 & 1 & 1 & 1 & 1 \\
AI/technology exposure at work & 2 & 8 & 7 & 7 & 12 \\
Life satisfaction & 3 & 2 & 3 & 2 & 2 \\
Trust in national institutions & 4 & 3 & 2 & 4 & 3 \\
Internet use & 5 & 5 & 10 & 3 & 10 \\
Age & 6 & 4 & 8 & 5 & 8 \\
Political interest & 7 & 10 & 4 & 11 & 6 \\
Occupation & 8 & 6 & 5 & 9 & 4 \\
Gender & 9 & 9 & 16 & 10 & 18 \\
Education & 10 & 11 & 9 & 8 & 5 \\
Community type & 11 & 13 & 12 & 14 & 9 \\
Sector of activity & 12 & 12 & 11 & 12 & 11 \\
Employment status & 13 & 17 & 14 & 17 & 16 \\
Foreign background & 14 & 16 & 15 & 13 & 15 \\
Workplace size & 15 & 15 & 13 & 18 & 13 \\
Household size & 16 & 14 & 17 & 16 & 14 \\
Children in household & 17 & 18 & 18 & 15 & 17 \\
Financial condition & 18 & 7 & 6 & 6 & 7 \\
\bottomrule
\end{tabular}

\vspace{0.12cm}

\begin{minipage}{0.96\textwidth}
\footnotesize
\textit{Notes:} Rank 1 indicates the predictor block producing the largest increase in cross-validated MSE when omitted. Rankings are computed separately for the observed outcome and each LLM-generated outcome using the same block-ablation procedure and respondent covariates. LLM outcomes refer to \(T=0.5\).
\end{minipage}
\end{table}

\subsection*{F.2 Effects}

For \emph{Effects}, the observed ranking is again led by trust in EU institutions and AI/technology exposure at work, followed by trust in national institutions. Claude reproduces these first three positions exactly, while Gemini retains all three blocks among its four highest-ranked predictors. GPT assigns the highest rank to national institutional trust, whereas Grok places AI/technology exposure considerably lower.

\begin{table}[H]
\centering
\scriptsize
\caption{Complete predictor rankings for the \emph{Effects} outcome}
\label{tab:ranking_effects}
\renewcommand{\arraystretch}{1.08}
\setlength{\tabcolsep}{5pt}
\begin{tabular}{lccccc}
\toprule
\textbf{Predictor block} & \textbf{Observed} & \textbf{Claude} & \textbf{GPT} & \textbf{Gemini} & \textbf{Grok} \\
\midrule
Trust in EU institutions & 1 & 1 & 2 & 1 & 1 \\
AI/technology exposure at work & 2 & 2 & 3 & 3 & 10 \\
Trust in national institutions & 3 & 3 & 1 & 4 & 2 \\
Gender & 4 & 7 & 17 & 10 & 11 \\
Life satisfaction & 5 & 4 & 6 & 2 & 3 \\
Political interest & 6 & 9 & 5 & 12 & 5 \\
Age & 7 & 6 & 8 & 5 & 8 \\
Occupation & 8 & 5 & 4 & 6 & 4 \\
Education & 9 & 11 & 9 & 8 & 6 \\
Community type & 10 & 12 & 12 & 14 & 9 \\
Workplace size & 11 & 14 & 13 & 13 & 15 \\
Internet use & 12 & 13 & 11 & 7 & 13 \\
Sector of activity & 13 & 10 & 10 & 11 & 12 \\
Foreign background & 14 & 16 & 18 & 15 & 18 \\
Employment status & 15 & 17 & 15 & 16 & 17 \\
Financial condition & 16 & 8 & 7 & 9 & 7 \\
Household size & 17 & 18 & 16 & 18 & 14 \\
Children in household & 18 & 15 & 14 & 17 & 16 \\
\bottomrule
\end{tabular}

\vspace{0.12cm}

\begin{minipage}{0.96\textwidth}
\footnotesize
\textit{Notes:} Rank 1 indicates the predictor block producing the largest increase in cross-validated MSE when omitted. Rankings are computed separately for the observed outcome and each LLM-generated outcome using the same block-ablation procedure and respondent covariates. LLM outcomes refer to \(T=0.5\).
\end{minipage}
\end{table}

\subsection*{F.3 Use}

The \emph{Use} outcome produces somewhat greater differences in ranking structure. AI and technology exposure at work is the leading predictor in the observed data, followed by trust in EU and national institutions. All four LLM-generated outcomes retain the two institutional-trust blocks near the top of the ranking, but direct workplace exposure is reproduced less consistently. Claude ranks it fourth, GPT second, Gemini fourth, and Grok seventh. Occupation also becomes more prominent in the synthetic rankings than in the observed ranking.

\begin{table}[H]
\centering
\scriptsize
\caption{Complete predictor rankings for the \emph{Use} outcome}
\label{tab:ranking_use}
\renewcommand{\arraystretch}{1.08}
\setlength{\tabcolsep}{5pt}
\begin{tabular}{lccccc}
\toprule
\textbf{Predictor block} & \textbf{Observed} & \textbf{Claude} & \textbf{GPT} & \textbf{Gemini} & \textbf{Grok} \\
\midrule
AI/technology exposure at work & 1 & 4 & 2 & 4 & 7 \\
Trust in EU institutions & 2 & 2 & 3 & 1 & 1 \\
Trust in national institutions & 3 & 1 & 1 & 2 & 2 \\
Age & 4 & 6 & 8 & 10 & 10 \\
Life satisfaction & 5 & 5 & 7 & 5 & 4 \\
Gender & 6 & 7 & 14 & 11 & 12 \\
Community type & 7 & 14 & 16 & 13 & 9 \\
Political interest & 8 & 11 & 5 & 15 & 5 \\
Occupation & 9 & 3 & 4 & 3 & 3 \\
Internet use & 10 & 10 & 11 & 7 & 13 \\
Workplace size & 11 & 12 & 12 & 9 & 14 \\
Employment status & 12 & 17 & 17 & 17 & 18 \\
Financial condition & 13 & 8 & 6 & 6 & 8 \\
Children in household & 14 & 18 & 13 & 18 & 17 \\
Foreign background & 15 & 16 & 18 & 16 & 16 \\
Education & 16 & 13 & 9 & 12 & 6 \\
Household size & 17 & 15 & 15 & 14 & 15 \\
Sector of activity & 18 & 9 & 10 & 8 & 11 \\
\bottomrule
\end{tabular}

\vspace{0.12cm}

\begin{minipage}{0.96\textwidth}
\footnotesize
\textit{Notes:} Rank 1 indicates the predictor block producing the largest increase in cross-validated MSE when omitted. Rankings are computed separately for the observed outcome and each LLM-generated outcome using the same block-ablation procedure and respondent covariates. LLM outcomes refer to \(T=0.5\).
\end{minipage}
\end{table}

The complete rankings confirm the summary measures reported in the main paper. Similarity is strongest for the broad \emph{General} and \emph{Effects} outcomes, particularly for Claude and Gemini, while \emph{Use} exhibits larger disagreements over the relative importance of specific predictors. At the same time, the principal institutional-trust blocks remain prominent across most synthetic outcomes. The detailed rankings therefore support the interpretation of substantive validity as partial rather than exact preservation of the observed predictor structure.

\end{document}